\documentclass[useAMS,usenatbib]{mn2e}

\usepackage[active]{srcltx}
\usepackage{graphicx}
\usepackage{txfonts}
\usepackage{natbib}
\usepackage{multirow}
\usepackage{longtable}
\usepackage[applemac]{inputenc}
\usepackage{rotating}
\usepackage{bigdelim}
\usepackage{subfig}

\usepackage{color}
\usepackage{soul}
\definecolor{jaune}{rgb}{1.0, 1.0, 0.0}

\definecolor{vero}{rgb}{1.0, 0.50, 0.75}

\newcommand{\kms}{km\,s$^{-1}\,$}
\newcommand{\vsini}{$v\sin i\,$}

\newcommand{\bz}{\ensuremath{\langle B_z\rangle}}
\newcommand{\nz}{\ensuremath{\langle N_z\rangle}}

\newcommand{\cpd}{CPD~$-$28$^{\circ}$~2561}
\def\gtrsim{\mathrel{\hbox{\rlap{\hbox{\lower4pt\hbox{$\sim$}}}\hbox{$>$}}}}
\def\ltsim{\mathrel{\hbox{\rlap{\hbox{\lower4pt\hbox{$\sim$}}}\hbox{$<$}}}}

\title[Rotation, spectral variability, magnetic geometry and magnetosphere of the Of?p star \cpd]{Rotation, spectral variability, magnetic geometry and magnetosphere of the Of?p star \cpd\thanks{Based on observations obtained at the Canada-France-Hawaii Telescope (CFHT) which is operated by the National Research Council of Canada, the Institut National des Sciences de l'Univers (INSU) of the Centre National de la Recherche Scientifique of France, and the University of Hawaii.}}
\author[G.A. Wade et al.]{G.A. Wade\thanks{E-mail: wade-g@rmc.ca}$^1$, R.H. Barb\'a$^2$, J. Grunhut$^3$, F. Martins$^4$, V. Petit$^5$\thanks{Currently at Dept. of Physics \& Space Sciences, Florida Institute of Technology, Melbourne, FL, USA}, J.O. Sundqvist$^6$,  
\newauthor{R.H.D. Townsend$^7$, N.R. Walborn$^8$, E. Alecian$^9$, E.J. Alfaro$^{10}$, J. Ma\' iz Apell\' aniz$^{10}$\thanks{Currently at Centro de Astrobiolog{\'\i}a, INTA-CSIC, Madrid, Spain.}, J.I. Arias$^2$,  }
\newauthor{R. Gamen$^{11}$, N. Morrell$^{12}$, Y. Naz\'e$^{13,14}$\thanks{FNRS Research Associate}, A. Sota$^{10}$, A. ud-Doula $^{15}$ and the MiMeS Collaboration}
\\
$^{1}$Department of Physics, Royal Military College of Canada, PO Box 17000 Station Forces, Kingston, ON, Canada K7K 7B4 \\
$^{2}$Departamento de F\'isica, Universidad de La Serena, Av. Cisternas 1200 Norte, La Serena, Chile\\
$^{3}$European Southern Observatories, Karl-Schwarzschild-Str. 2, 85748, Garching, Germany\\
$^{4}$LUPM-UMR5299, CNRS \& Universit\'e Montpellier II, Place Eug\`ene Bataillon, F-34095, Montpellier, France\\
$^5$Department of Physics and Astronomy, Bartol Research Institute, University of Delaware, Newark, DE 19716, USA\\
$^6$Institut f\"ur Astronomie und Astrophysik der Universit\"at M\"unchen, Scheinerstr. 1, D-81679 M\"unchen, Germany\\
$^7$Department of Astronomy, University of Wisconsin-Madison, 475 N Charter Street, Madison, WI 53706, USA\\
$^8$Space Telescope Science Institute, 3700 San Martin Drive, Baltimore, MD 21218, USA\\
$^9$UJF-Grenoble 1/CNRS-INSU, Institut de PlanÈtologie et d'Astrophysique de Grenoble (IPAG) UMR 5274, 38041, Grenoble, France\\
$^{10}$Instituto de Astrof\'isica de Andaluc\' ia-CSIC, Glorieta de la Astronom\'ia s/n, E-18008 Granada, Spain\\
$^{11}$Instituto de Astrof\'isica de La Plata (CCT La Plata-CONICET, Universidad Nacional de La Plata), Paseo del Bosque s/n, 1900 La Plata, Argentina\\
$^{12}$Las Campanas Observatory, Observatories of the Carnegie Institution of Washington, La Serena, Chile\\
$^{13}$FNRS-GAPHE, D\'epartement AGO, Universit\'e de Li\`ege, All\'ee du 6 Ao\^ut 17, Bat. B5C, B4000-Li\`ege, Belgium\\
$^{14}$Groupe d'Astrophysique des Hautes Energies, Institut d'Astrophysique et de G\'eophysique, Universit\'e de Li\`ege, 17, All\'ee du 6 Ao\^ut, B5c, B-4000 Sart Tilman, Belgium\\
$^{15}$Penn State Worthington Scranton, Dunmore, PA 18512, USA}

\begin{document}

\date{Accepted . Received , in original form }

\pagerange{\pageref{firstpage}--\pageref{lastpage}} \pubyear{2002}

\maketitle

\label{firstpage}

\begin{abstract}

We report magnetic and spectroscopic observations and modeling of the Of?p star \cpd. Using more than 75 new spectra, we have measured the equivalent width variations and examined the dynamic spectra of photospheric and wind-sensitive spectral lines. A period search results in an unambiguous $73.41$~d variability period. High resolution spectropolarimetric data analyzed using Least-Squares Deconvolution yield a Zeeman signature detected in the mean Stokes $V$ profile corresponding to phase 0.5 of the spectral ephemeris. Interpreting the 73.41~d period as the stellar rotational period, we have phased the equivalent widths and inferred longitudinal field measurements. The phased magnetic data exhibit a weak sinusoidal variation, with maximum of about 565 G at phase 0.5, and a minimum of about -335 G at phase 0.0, with extrema approximately in phase with the (double-wave) H$\alpha$ equivalent width variation. Modeling of the H$\alpha$ equivalent width variation assuming a quasi-3D magnetospheric model produces a unique solution for the ambiguous couplet of inclination and magnetic obliquity angles: $(i, \beta)$ or $(\beta, i)=(35\degr,90\degr)$. Adopting either geometry, the longitudinal field variation yields a dipole polar intensity $B_{\rm d}=2.6\pm 0.9$~kG, consistent with that obtained from direct modelling of the Stokes $V$ profiles. We derive a wind magnetic confinement parameter $\eta_*\simeq 100$, leading to an Alfv\'en radius $R_{\rm A}\simeq 3-5~R_*$, and a Kepler radius $R_{\rm K}\simeq 20~R_*$. This supports a physical scenario in which the H$\alpha$ emission and other line variability have their origin in an oblique, co-rotating 'dynamical magnetosphere' structure resulting from a magnetically channeled wind. Nevertheless, the details of the formation of spectral lines and their variability within this framework remain generally poorly understood. 
%In particular, the origin of the periodic velocity variations observed in emission lines of \cpd\ and one other star is an outstanding problem. 
%Thirty-three low and high resolution spectroscopic observations have been obtained in the context of the O and WN-type Stars (OWN) survey and the Galactic O Stars Spectroscopic Survey (GOSSS). In addition, 44 high resolution circularly polarised (Stokes $V$) spectra were obtained in the context of the Magnetism in Massive Stars (MiMeS) project. 
 \end{abstract}


% ========================================================================
% ========================================================================

\section{Introduction}

The Of?p stars are mid-O-type stars identified by a number of peculiar observational properties. The classification was first introduced by \citet{1972AJ.....77..312W} according to the presence of C~{\sc iii} $\lambda 4650$ emission with a strength comparable to the neighbouring N~{\sc iii} lines. Well-studied Of?p stars are now known to exhibit recurrent, and apparently periodic, spectral variations (in Balmer, He~{\sc i}, C~{\sc iii} and Si~{\sc iii} lines), narrow P Cygni or emission components in the Balmer lines and He~{\sc i} lines, and UV wind lines weaker than those of typical Of supergiants \citep[see][and references therein]{2010A&A...520A..59N}. 

Only 5 Galactic Of?p stars are known \citep{Walbetal10a}: HD 108, HD 148937, HD 191612, NGC 1624-2 and \cpd. Four of these stars - HD 108, HD 148937, HD 191612 and NGC 1624-2 - have been studied in detail based on optical spectra, and in some cases UV spectra as well. In recent years, they have been carefully examined for the presence of magnetic fields \citep{2006MNRAS.365L...6D,2010MNRAS.407.1423M,2011MNRAS.416.3160W,2012MNRAS.419.2459W,2012MNRAS.425.1278W} and all have been clearly detected. It therefore appears that the particular spectral peculiarities that define the Of?p classification are a consequence of their magnetism. Indeed, \citet{2011A&A...528A.151H,2012IBVS.6019....1H} reported 4 FORS+VLT measurements of the longitudinal magnetic field of \cpd\ at the few hundred G level, several of which correspond to detections at somewhat more than 3$\sigma$ significance. 

Like HD 108, HD 191612, HD~148937 and NGC 1624-2, \cpd\ is a spectroscopic variable star \citep[e.g.][]{Walbetal10a}. The spectrum of \cpd\ was first described as a peculiar Of type by \citet{1973AJ.....78.1067W} and \citet{1977ApJS...35..111G}. Garrison et al. commented: 'Very peculiar spectrum.  Carbon (C~{\sc iii} $\lambda$4070) is strong, nitrogen weak.  H and He~{\sc ii} lines are broad, while He~{\sc i} lines are sharp,'  Walborn classified the spectrum as O6.5fp.  The outstanding peculiarity was that although He~{\sc ii} $\lambda$4686 emission was very strong, appropriate for an Of supergiant, the N~{\sc iii} $\lambda$4640 emission was incompatibly not.

%; the latter authors kindly permitted a review of their survey material by the former, who subsequently reobserved it at somewhat higher resolution.  

% without explanation in deference to the forthcoming survey paper

%where the remark about carbon refers to the C~{\sc iii} $\lambda$4070 absorption blend.  

% It was already classified as a peculiar Of object of undetermined nature by Walborn (1973a) and Garrison, Hiltner, \& Schild (1977). Walborn et al. (2010) mentioned that repeated high-resolution observations in the context of the GOSSS show that this star undergoes extreme spectral transformations very similar to those of HD 191612, on timescales of weeks. 

While \citet{1973AJ.....78.1067W}  announced HD~191612 as a new, third member of the Of?p class, it is significant that no such association was made for
\cpd.  The reason was that no C~{\sc iii} $\lambda$4650 emission was detected in \cpd, while a comparable emission strength to that of N~{\sc iii} was the primary defining Of?p characteristic. (Of course, it was subsequently discovered that the C~{\sc iii} emission disappears entirely at the minimum phase of HD~191612 \citep{2004ApJ...617L..61W, 2007MNRAS.381..433H,2011MNRAS.416.3160W}).  It was not until the intensive OWN survey's high-resolution monitoring of \cpd\ \citep{2010RMxAC..38...30B} revealed extreme variations in $\lambda$4686 and Balmer lines, analogous to those of HD~191612, that the association was made \citep{Walbetal10a} and \cpd\ identified as an Of?p star.

%In retrospect, the peculiarities noted in the early studies are also characteristic of the Of?p class.  However, its definition must now be expanded to include {\it weak} C~III emission.  As further elaborated below, this diversity of detail among the Of?p class will likely contribute to eventual improved understanding of all of them.

In this paper we perform a first detailed investigation of the combined magnetic and variability properties of \cpd\ using an extensive spectroscopic and high-resolution spectropolarimetric dataset. In Sect. 2 we discuss the data acquired and the methods of analysis used. In Sect. 3 we re-examine the physical properties of the star, as well as its projected rotational velocity.  In Sect. 4 we examine the spectral characteristics and variability, identifying periodic variability of the H$\alpha$ and other emission and absorption lines and deriving the rotational period of the star. In Sect. 5 we analyse in detail the magnetic data acquired at the Canada-France-Hawaii Telescope (CFHT). In Sect. 6 we employ the H$\alpha$ EW variation and CFHT magnetic data to constrain the stellar and magnetic geometry and the surface field strength. In Sect. 7 we derive the magnetospheric properties of \cpd. Finally, in Sects. 8 we summarize our results, and explore the implications of our study, particularly regarding the variability and other properties of \cpd, the confinement and structure of its stellar wind, and of the properties of the general class of Of?p stars.

%\begin{figure*}
%\begin{centering}
%\includegraphics[width=14cm,angle=-90]{CPD-282561_BzEW.ps}
%%\vspace{3in}
%\caption{\label{}EW and \bz\ variability}
%\end{centering}
%\end{figure*} 

\section{Observations}

\begin{table}
\caption{Spectrographs used for the acquisition of \cpd\ spectroscopy and spectropolarimetry.}
\label{spectrographs}
\begin{center}
\begin{tabular}{rrcrrrrr}
\hline
Spectrograph    &    Telescope     &        Spectral Range   &Resolving \\
 & & (\AA) & Power\\
 \hline
Echelle      &       2.5  m LCO/du Pont  &  3600 -- 9200  &   46,000\\
FEROS       &        2.2  m ESO/MPI      &  3500 -- 9200    & 48,000\\
REOSC        &       2.15 m CASLEO      &   3800 -- 6000   &  15,000\\
Boller \& Chivens &  2.5  m LCO/du Pont &   3900 -- 5500  &    2,500\\
ESPaDOnS & 3.6 m CFHT & 3650 -- 10000 & 65,000\\
\hline
\end{tabular}
\end{center}
\end{table}

\begin{table}
\caption{Log of spectroscopic observations showing heliocentric Julian Date, rotational phase according to Eq. (1), and exposure time. The column 'Spectrograph' corresponds to the facilities described in Table 1.}
\label{table:spectroscopy}
\begin{center}
\begin{tabular}{lcrrrrrrrr}
\hline
HJD      &     Phase & Exp. time  & Spectrograph  \\
         &           & (s)        &           \\
\hline
  2453875.483    &       0.511 &   600   &  Echelle  \\
 2454246.515    &       0.565 & 1800    & FEROS   \\
 2454610.486    &       0.523 &  900    & B\&C    \\
 2454627.467    &       0.754 & 1800    & FEROS    \\
 2454786.865    &       0.926 &  900    & Echelle   \\
 2454842.716    &       0.687 &  1800   &  REOSC  \\
 2454846.677    &       0.741 &  1800   &  REOSC   \\
 2454847.609    &       0.753 &  2000   &  REOSC   \\
 2454848.650    &       0.767 &  1800   &  REOSC   \\
 2454954.537    &       0.210 &  1800   &  FEROS    \\
 2454955.500    &       0.223 &  1800   &  FEROS    \\
 2454964.476    &       0.345 &  1500   &  Echelle  \\
 2455341.455    &       0.481 &  1200   &  Echelle  \\
 2455495.865    &       0.584 &   900   &  B\&C     \\
 2455604.796    &       0.068 &  1800   &  REOSC    \\
 2455641.586    &       0.569 &  1800   &  REOSC    \\
 2455643.569    &       0.596 &  1800   &  REOSC    \\
 2455646.597    &       0.637 &   900   &  B\&C     \\
 2455669.553    &       0.950 &   900   &  B\&C     \\
 2455672.532    &       0.990 &  1200   &  Echelle  \\
 2455696.473    &       0.317 &  2700   &  REOSC    \\
 2455697.494    &       0.331 &  2400   &  REOSC    \\
 2455698.518    &       0.344 &  2400   &  REOSC    \\
 2455699.508    &       0.358 &  2400   &  REOSC    \\
 2455716.479    &       0.589 &   900   &  B\&C     \\
 2455898.729    &       0.072 &   900   &  B\&C     \\
 2455976.599    &       0.133 &  1200   &  Echelle  \\
 2456053.500    &       0.180 &   900   &  B\&C     \\
 2456080.469    &       0.547 &  1800   &  Echelle  \\
 2456098.461    &       0.793 &  2000   &  FEROS    \\
 2456340.705    &       0.092 &   900   &  B\&C     \\
 2456367.653    &       0.460 &   900   &  B\&C     \\
 2456381.587    &       0.649 &   900   &  B\&C     \\
 \hline\hline
\end{tabular}
\end{center}
\end{table}

\subsection{Spectroscopic observations}

Spectroscopic observations of \cpd\ were obtained in the 
framework of two spectroscopic surveys: the {High-resolution spectroscopic
monitoring of Southern  O and WN-type Stars} \citep[The "OWN Survey",][]{2010RMxAC..38...30B} and the {Galactic O Star Spectroscopic Survey} \citep["GOSSS",][]{2011hsa6.conf..467M}
In the OWN Survey program,  high-resolution and high
signal-to-noise 
spectra are being acquired for a sample of 240 massive southern stars selected from the {
Galactic O Star Catalogue} \citep[GOSC version 1,][]{2004ApJS..151..103M}. One of the goals is to determine precise radial velocities in order
to detect new binaries among these stars for which there is scarce or no
indication of multiplicity. An additional goal is to detect possible
spectral variations which can be related to multiplicity, the presence of
magnetic fields, pulsation, or eruptive behaviour. 
\cpd\ was included early in the observed sample, and monitored
systematically as large variations in the intensity of the He~{\sc ii}
$\lambda 4686$ emission line were detected. The star was observed spectroscopically from
three different locations: Las Campanas Observatory (LCO), and La Silla
Observatory (LSO), both in Chile, and Complejo Astron\'omico El Leoncito
(CASLEO), in Argentina, during 33 nights between 2006 and 2012. 

Seven spectrograms were obtained with an \'echelle spectrograph attached to the
2.5 m LCO/du Pont telescope. The spectral resolving power is about 46,000. 
Twelve spectrograms were obtained with the FEROS \'echelle spectrograph
attached to the 
2.2~m ESO/MPI telescope. In this case, the spectral resolving power is about
48,000.  
Additionally, four spectrograms were obtained with the REOSC \'echelle
spectrograph\footnote{Jointly built by REOSC and Li\`ege Observatory, and on long-term loan from the latter.} attached to the 2.15 m CASLEO/Jorge Sahade telescope, with a
resolving power of about 15,000.
Thorium-Argon comparison lamp exposures were obtained before or after the science
exposures. 
LCO and CASLEO \'echelle spectrograms were reduced using the {\sc Echelle} package
layered in IRAF\footnote{IRAF is the Image Reduction and Analysis Facility, a general purpose software system for the reduction and analysis of astronomical data. IRAF is written and supported by the National Optical Astronomy Observatories (NOAO) in Tucson, Arizona. See {\tt iraf.noao.edu}.}, while FEROS spectrograms were reduced
using the standard MIDAS pipeline. All spectrograms were bias subtracted,
flat-fielded, echelle order identified, and extracted, and finally wavelength
calibrated. Table~\ref{spectrographs} presents technical details of the
different spectrographs utilized.  

Under the GOSSS program, we have obtained ten spectrograms of  \cpd\
between 2008 and 2013 using the Boller \& Chivens spectrograph attached
to the 2.5 m LCO/du Pont telescope, with a spectral resolving power of about
2,500. Helium-Neon-Argon comparison lamps were used for wavelength
calibration. A dedicated pipeline was developed for the complete reduction of
GOSSS observations. Detailed description about the observing
procedures and data reduction are described by Sota et al. (2011) and Sota et
al. (2014). 

The log of spectroscopic observations is reported in Table~\ref{table:spectroscopy}.

\subsection{Spectropolarimetric observations}

Spectropolarimetric observations of \cpd\ were obtained using the ESPaDOnS spectropolarimeter at the CFHT in 2012 and 2013 within the context of the Magnetism in Massive Stars (MiMeS) Large Program \citep[][Wade et al., in prep.]{2014IAUS..302..265W}. Altogether, 44 Stokes $V$ sequences were obtained.  

Each polarimetric sequence consisted of four individual subexposures taken in different polarimeter configurations. From each set of four subexposures we derive a mean Stokes $V$ spectrum following the procedure of \citet{1997MNRAS.291..658D}, ensuring in particular that all spurious signatures are { suppressed} at first order. Diagnostic null polarization spectra (labeled $N$) are calculated by combining the four subexposures in such a way that polarization cancels out, allowing us to check that no spurious signals are present in the data (see \citet{1997MNRAS.291..658D} for more details on how $N$ is defined). All frames were processed using the Upena pipeline feeding Libre ESpRIT \citep{1997MNRAS.291..658D}, a fully automatic reduction package installed at CFHT for optimal extraction of ESPaDOnS spectra. The peak signal-to-noise ratios per 1.8 km\,s$^{-1}$ velocity bin in the reduced spectra range from about 70 to nearly 266, with a median of 200, depending on the exposure time and on weather conditions. 

The log of CFHT observations is presented in Table~\ref{tab:specpol}.

{ We also obtained one observation of \cpd\ with the HARPS instrument in polarimetric model. The observing procedure was fundamentally the same as that described above for ESPaDOnS. Two consecutive observations were acquired with exposure times of 3600s apiece. The peak SNR of the combined spectrum was 110 per 1~km/s spectral pixel. The mean HJD of the observation is 2455905.798, corresponding to phase 0.168 according to the ephemeris defined in Eq. (1).}

% Table is up to date acc to directory [muffin-air:CPD-282561/analysis/ilsd_hd191612mask_maxabs20] wade 3 June 2014
% Phases updated using new ephemeris.

\begin{table*}
\caption{Log of ESPaDOnS spectropolarimetric observations. Columns indicate CFHT archive unique identifier, heliocentric Julian Date, exposure duration, signal-to-noise ratio per 1.8~km/s pixel, phase according to Eq. (1), mean phase of indicated binned spectra, Stokes $V$ detection probability, longitudinal magnetic field and longitudinal field significance, diagnostic null detection probability, longitudinal magnetic field and longitudinal field significance.}
\label{tab:specpol}
\begin{center}
\begin{tabular}{cccrrp{-0.1cm}rrrrrrrrr}
Odometer        &       HJD     &       Exp     &       SNR     &       Phase   &               &       $\overline{\rm Phase}$                  &       $P_{\rm V}$     &       $\bz$   &       $|z_V|$ &       $P_{\rm N}$     &       $\nz$   &       $|z_N|$ \\
        &               &       (s)     &       (/pix)  &               &               &                               &       (\%)    &               &               &       (\%)    &               &               \\
\hline                                                                                                                                                                                                                  
1051146 &       2454844.994     &       4340    &       204     &       0.718&          &                               &               &               &               &               &               &               \\
1051150 &       2454845.047     &       4340    &       204     &       0.718&          &       \multirow{-2}{*}{       0.718   }       &       \multirow{-2}{*}{18.179}        &       \multirow{-2}{*}{$73\pm 275$}   &       \multirow{-2}{*}{0.3}   &       \multirow{-2}{*}{4.580} &       \multirow{-2}{*}{$+128\pm 274$} &       \multirow{-2}{*}{0.5}   \\
1513790 &       2455931.006     &       4340    &       264     &       0.511&          &                               &               &               &               &               &               &               \\
1513798 &       2455931.084     &       4340    &       244     &       0.513&          &                               &               &               &               &               &               &               \\
1513935 &       2455931.861     &       4340    &       235     &       0.523&          &                               &               &               &               &               &               &               \\
1513939 &       2455931.913     &       4340    &       238     &       0.524&          &                               &               &               &               &               &               &               \\
1514254 &       2455933.035     &       4340    &       235     &       0.539&          &                               &               &               &               &               &               &               \\
1514415 &       2455934.004     &       4340    &       211     &       0.552&   \ldelim\} {-6}{0.1in}  &       \multirow{-6}{*}{       0.527   }       &       \multirow{-6}{*}{89.468}        &       \multirow{-6}{*}{$-253\pm 174$} &       \multirow{-6}{*}{1.5}   &       \multirow{-6}{*}{26.656}        &       \multirow{-6}{*}{$-268\pm 173$} &       \multirow{-6}{*}{1.6}   \\
1514600 &       2455935.001     &       4340    &       266     &       0.566&          &                               &               &               &               &               &               &               \\
1514818 &       2455936.014     &       4340    &       243     &       0.580&          &                               &               &               &               &               &               &               \\
1515076 &       2455936.952     &       4340    &       252     &       0.592&          &                               &               &               &               &               &               &               \\
1515397 &       2455938.928     &       4340    &       263     &       0.619&   \ldelim\} {-4}{0.1in}  &       \multirow{-4}{*}{       0.589   }       &       \multirow{-4}{*}{21.516}        &       \multirow{-4}{*}{$-362\pm 186$} &       \multirow{-4}{*}{2.0}   &       \multirow{-4}{*}{68.539}        &       \multirow{-4}{*}{$+144\pm 186$} &       \multirow{-4}{*}{0.8}   \\
1515606 &       2455940.963     &       4340    &       191     &       0.647&          &                               &               &               &               &               &               &               \\
1516007 &       2455942.995     &       4340    &       265     &       0.675&          &                               &               &               &               &               &               &               \\
1516011 &       2455943.048     &       4340    &       241     &       0.675&          &                               &               &               &               &               &               &               \\
1516165 &       2455943.988     &       4340    &       259     &       0.688&   \ldelim\} {-4}{0.1in}  &       \multirow{-4}{*}{       0.671   }       &       \multirow{-4}{*}{25.299}        &       \multirow{-4}{*}{$+300\pm 170$} &       \multirow{-4}{*}{1.8}   &       \multirow{-4}{*}{0.028} &       \multirow{-4}{*}{$-104\pm 170$} &       \multirow{-4}{*}{0.6}   \\
1524346 &       2455967.882     &       4340    &       212     &       0.014&          &                               &               &               &               &               &               &               \\
1524350 &       2455967.936     &       4340    &       209     &       0.015&          &                               &               &               &               &               &               &               \\
1524524 &       2455968.943     &       4340    &       204     &       0.028&          &                               &               &               &               &               &               &               \\
1524528 &       2455968.997     &       4340    &       178     &       0.029&          &                               &               &               &               &               &               &               \\
1524741 &       2455969.865     &       4340    &       233     &       0.041&          &                               &               &               &               &               &               &               \\
1524745 &       2455969.918     &       4340    &       247     &       0.042&   \ldelim\} {-6}{0.1in}  &       \multirow{-6}{*}{       0.028   }       &       \multirow{-6}{*}{68.229}        &       \multirow{-6}{*}{$+422\pm 223$} &       \multirow{-6}{*}{1.9}   &       \multirow{-6}{*}{11.331}        &       \multirow{-6}{*}{$+246\pm 223$} &       \multirow{-6}{*}{1.1}   \\
1597878 &       2456259.116     &       3000    &       188     &       0.981&          &                               &               &               &               &               &               &               \\
1597882 &       2456259.153     &       3000    &       168     &       0.982&         &    \multirow{-2}{*}{   0.981   }       &       \multirow{-2}{*}{92.015}        &       \multirow{-2}{*}{$+650\pm 414$}  &       \multirow{-2}{*}{1.6}   &       \multirow{-2}{*}{15.574}        &       \multirow{-2}{*}{$+300\pm 415$} &       \multirow{-2}{*}{0.7}   \\
1601442 &       2456283.041     &       3000    &       127     &       0.307&          &                               &               &               &               &               &               &               \\
1601446 &       2456283.079     &       3000    &       128     &       0.307&          &                               &               &               &               &               &               &               \\
1601450 &       2456283.116     &       3000    &       145     &       0.308&          &                               &               &               &               &               &               &               \\
1601454 &       2456283.153     &       3000    &       138     &       0.308&          &                               &               &               &               &               &               &               \\
1602166 &       2456285.940     &       3000    &       200     &       0.346&          &                               &               &               &               &               &               &               \\
1602170 &       2456285.977     &       3000    &       200     &       0.347&          &                               &               &               &               &               &               &               \\
1602174 &       2456286.013     &       3000    &       205     &       0.347&          &                               &               &               &               &               &               &               \\
1602178 &       2456286.050     &       3000    &       198     &       0.348&   \ldelim\} {-8}{0.1in}  &       \multirow{-8}{*}{       0.327   }       &       \multirow{-8}{*}{35.563}        &       \multirow{-8}{*}{$-243\pm 168$} &       \multirow{-8}{*}{1.5}   &       \multirow{-8}{*}{0.848} &       \multirow{-8}{*}{$+51\pm 169$}  &       \multirow{-8}{*}{0.3}   \\
1603502 &       2456290.985     &       3000    &       139     &       0.415&          &                               &               &               &               &               &               &               \\
1603506 &       2456291.022     &       3000    &       131     &       0.416&          &                               &               &               &               &               &               &               \\
1603510 &       2456291.059     &       3000    &       139     &       0.416&          &                               &               &               &               &               &               &               \\
1603514 &       2456291.095     &       3000    &       138     &       0.417&   \ldelim\} {-4}{0.1in}  &       \multirow{-4}{*}{       0.416   }       &       \multirow{-4}{*}{96.791}        &       \multirow{-4}{*}{$-398\pm 361$} &       \multirow{-4}{*}{1.1}   &       \multirow{-4}{*}{27.388}        &       \multirow{-4}{*}{$-518\pm 359$} &       \multirow{-4}{*}{1.4}   \\
1604359 &       2456294.875     &       3000    &       142     &       0.468&          &                               &               &               &               &               &               &               \\
1604363 &       2456294.912     &       3000    &       111     &       0.469&          &                               &               &               &               &               &               &               \\
1604367 &       2456294.949     &       3000    &       107     &       0.469&          &                               &               &               &               &               &               &               \\
1604371 &       2456294.986     &       3000    &       105     &       0.470&          &                               &               &               &               &               &               &               \\
1604375 &       2456295.023     &       3000    &       102     &       0.470&          &                               &               &               &               &               &               &               \\
1604379 &       2456295.060     &       3000    &       67      &       0.471&          &                               &               &               &               &               &               &               \\
1604383 &       2456295.097     &       3000    &       73      &       0.471&          &                               &               &               &               &               &               &               \\
1604387 &       2456295.134     &       3000    &       76      &       0.472&   \ldelim\} {-8}{0.1in}  &       \multirow{-8}{*}{       0.470   }       &       \multirow{-8}{*}{99.978}        &       \multirow{-8}{*}{$+120\pm 361$} &       \multirow{-8}{*}{0.3}   &       \multirow{-8}{*}{84.634}        &       \multirow{-8}{*}{$-282\pm 362$} &       \multirow{-8}{*}{0.8}   \\
		  \hline
\end{tabular}
\end{center}
\end{table*}

\section{Stellar physical and wind properties}

\begin{table}
\centering
\caption{Summary of physical, wind and magnetic properties of \cpd. Apart from the luminosity $L/L_\odot$ (which was adopted), all parameters are derived in this paper. The mass-loss rate corresponds to the CMFGEN radiatively-driven rate required to reasonably reproduce the H$\alpha$ profile; this value is only indicative, since the spherical symmetry assumed by CMFGEN is clearly broken in the case of \cpd. The wind magnetic confinement parameter $\eta_*$, the rotation parameter $W$ and the characteristic spin down time $\tau_{\rm spin}$ are defined and described in Sect. 7. Values in brackets for $\eta_*$, $R_{\rm A}$ and $R_{\rm K}$ correspond to the best-fit parameters.}
\begin{tabular}{l|ll}
\hline
Spectral type & Of?p             \\
$T_{\rm eff}$ (K) & 35 000 $\pm$ 2000  \\
$\log g$ (cgs) & 4.0 $\pm$ 0.1     \\
R$_{\star}$ (R$_\odot$) & 12.9 $\pm 3.0$  \\
$\log (L_\star/L_\odot)$ & 5.35$\pm 0.15$ \\
$M_{\rm evol}$ ($M_{\odot}$) & 35$\pm$6  \\
$M_{\rm spec}$ ($M_{\odot}$) & 61$\pm$33 \\
\hline
$v\sin i$ (km\,s$^{-1}$) & $\ltsim 80$   \\
$P_{\rm rot}$ (d) & $73.41\pm 0.05$ \\
\hline
%\multicolumn{3}{c}{UV wind} \\
%$\log \dot{M}/\sqrt{f}$ (M$_{\odot}$\,yr$^{-1}$) & $-6.0$  \\
$\log \dot{M}_{\rm B=0}$ (M$_{\odot}$\,yr$^{-1}$) & $-6.0$  \\
$v_{\infty}$ (km\,s$^{-1}$) & 2400  \\
%Clumping $f$ & 0.1 & Assumed\\
%\hline
%\multicolumn{3}{c}{H$\alpha$ wind} \\
%$\log \dot{M}/\sqrt{f}$ (M$_{\odot}$\,yr$^{-1}$) & $-5.8$  \\
%$v_{\infty}$ (km\,s$^{-1}$) & 900  \\
\hline
$B_{\rm d}$ (G) & $2600\pm 900$ \\
$i$ ($\degr$) & $35\pm 3$ \\
$\beta$ ($\degr$) & $90\pm 4$ \\
\hline
$\eta_*$ & 20-900 (93)\\
$R_{\rm A}$ ($R_*$) & 2.1-5.5 (3.4)\\
%$W$ & \vero{$??\times 10^{??}$} \\
$R_{\rm K}$ ($R_*$) & 14.2-29.9 (18.7)\\
$\tau_{\rm spin}$ & 0.45 Myr \\
%\hline
%\multicolumn{3}{c}{Slow wind} \\
%$\log \dot{M}$ (M$_{\odot}$\,yr$^{-1}$) & $-5.8$  & This paper\\
%$v_{\infty}$ (km\,s$^{-1}$) & 900  & This paper\\
%Fill factor $f$ & Not measured\\
\hline
N/H & $<5\times$10$^{-5}$\\
C/H & $(5\pm 3)\times$10$^{-5}$ \\
\hline\hline
\end{tabular}
\label{params}
\end{table}

We have used the atmosphere code CMFGEN \citep{hm98} to determine the stellar parameters of \cpd. CMFGEN computes non-LTE, spherical models including an outflowing wind and line-blanketing. A complete description of the code is given by Hillier \& Miller (1998). We have included the following elements in our models: H, He, C, N, O, Ne, Mg, Si, S, Ar, Ca, Fe and Ni. A solar metallicity \citep{2010Ap&SS.328..179G} was adopted.

\begin{figure}
\hspace{-0.5cm}\includegraphics[width=9cm]{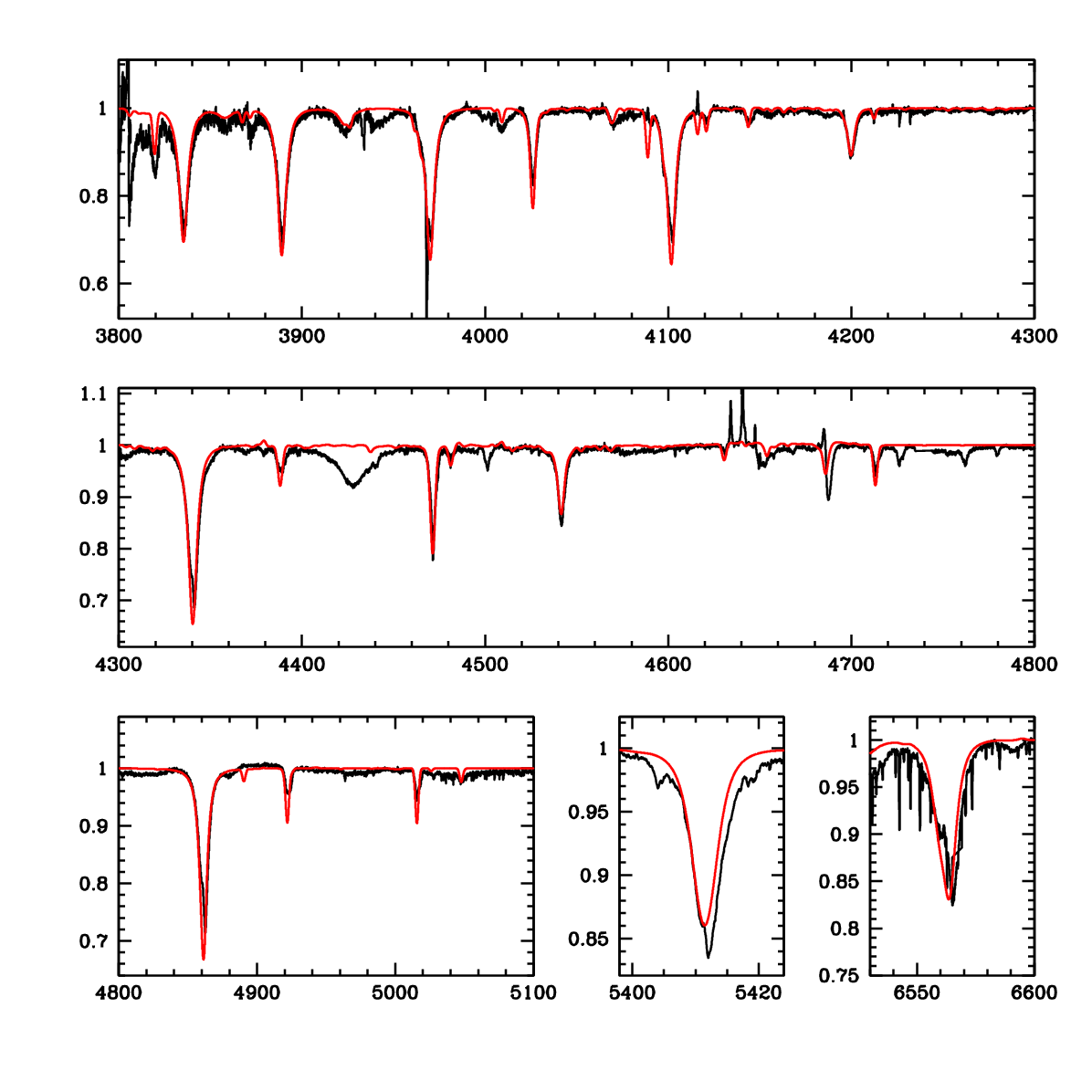}
\caption[]{Best fit CMFGEN model (red) of the observed ESPaDOnS spectrum (black, { phase 0.72, average of spectra \#s 1051146/50}) for \cpd.}
\label{fig_fit}
\end{figure}

We used the classical helium ionization balance method to constrain the effective temperature. As shown in Sect.\ 4, spectral variability is observed in He~{\sc i} lines. Part of the variability may be due to emission from { the stellar wind (i.e. the magnetosphere inferred later in the paper),} contaminating the underlying photospheric features. The determination of $T_{\rm eff}$ is thus difficult. It was performed using spectra { \#1051146/50 (corresponding to phase 0.718 of the rotational ephemeris described in Sect. 4)} where the { He~{\sc i} lines were the strongest, presumably minimizing contamination from the wind and} leaving access to the { cleanest} photospheric profiles. { Nevertheless, some wind-sensitive lines are poorly reproduced. Notably He~{\sc ii} $\lambda 4686$ exhibits a P Cyg-like profile, with a strong redshifted absorption that (unlike most other absorption features in the spectrum) is under-fit by the model. The profile of this line may be suggestive of accretion.} 

The relative strength of He~{\sc i} and He~{\sc ii} lines indicates a temperature of about 35000~K, with an uncertainty of 2000 K. The surface gravity was determined from the wings of the numerous Balmer lines available in the ESPaDOnS spectrum. The line cores are variable but are not the main gravity diagnostics so that our $\log g$ estimate is safe. We found that $\log g$ = 4.0$\pm$0.1 provided the best fit. In absence of strong constraint on the distance of \cpd, we decided to adopt a luminosity of $10^{5.35\pm0.15}$ $L_\odot$. This value is intermediate between that of an O6.5 dwarf and giant \citep{msh05}, and similar to that of the Of?p star HD 191612 \citep{2007MNRAS.381..433H}. The corresponding radius is thus 12.9$\pm$3.0 $R_{\odot}$. The best fit of the spectrum with maximum He~{\sc i}~4471 absorption is shown in Fig.\ \ref{fig_fit}. 

The evolutionary mass, determined from the position in the HR diagram and interpolation between the tracks of Meynet \& Maeder (2005), is 35$\pm$6 $M_{\odot}$. The spectroscopic mass, obtained from the surface gravity and the radius, is 61$\pm$33 $M_{\odot}$. The mass estimates are consistent within the error bars, which remain large because of the uncertainties on the distance (and thus on the luminosity). We also note that the evolutionary tracks adopted here do not include the effects of a strong, large-scale stellar magnetic field.

The shape of the photospheric lines could be reproduced with different combinations of rotational velocities and macroturbulence. In the extreme cases, a macroturbulent isotropic velocity of 40 \kms\ and a negligible rotational velocity, or no macroturbulence and \vsini\ = 80 \kms, correctly reproduce the shape of the photospheric lines. 
Using C~{\sc iii} and N~{\sc iii} lines as indicators (C~{\sc iii} $\lambda$4070 and N~{\sc iii} $\lambda$4505--4515 being the main diagnostics), we estimated the nitrogen and carbon content to be N/H$<$5.0$\times$10$^{-5}$ and C/H=(5.0$\pm$3.0)$\times$10$^{-5}$. Both elements are underabundant compared to the Sun, which seems consistent with the location of the star in the outer part of the disk, beyond the solar circle. The carbon to nitrogen ratio is consistent with little enrichment: N/C$<$1.0, compared to 0.25 for the sun \citep{2010Ap&SS.328..179G}. There is no evidence for strong He enrichment. Hence the abundances of \cpd\ have barely been affected by chemical processing occurring inside the star.

Finally, we adopted a wind terminal velocity of 2400 km s$^{-1}$, and a velocity field slope $\beta$=1.0. These values are typical of O dwarfs/giants. We found that a mass loss rate of 10$^{-6}$ M$_{\odot}$ yr$^{-1}$ leads to a density $\rho = {\dot M_{\rm B=0}}/(4 \pi r^2 v)$ able to correctly reproduce the shape of H${\alpha}$ in the spectrum showing the weakest nebular contamination. We caution that this value of $\dot M$ should not be blindly interpreted as the true mass loss rate of this magnetic star's outflow. As discussed in detail in Sect. 7, the circumstellar structure of \cpd\ is expected to be highly aspherical and dominated by infalling material. MHD models by ud-Doula et al. (2008) show that for a star like \cpd, the true rate of mass lost from the star into interstellar space (i.e. from the top of the magnetosphere) is reduced by a factor of about 5, i.e. $\sim 80$\% of the material leaving the surface actually falls back upon the star. Hence the true mass loss rate is most likely significantly lower than 10$^{-6}$ M$_{\odot}$ yr$^{-1}$. Rather, the CMFGEN $\dot M_{\rm B=0}$ should be viewed as a parameter giving a first rough estimate of the expected circumstellar density (see further Sect. 7).

%It's right in the plane.  SIMBAD gives some apparently significant proper motions...  Does this mean it is a wild runaway, or subluminous and more nearby...?  But the reddening is %considerable...  Just to muddy the water, you know, I was musing the other day that Ofp PNN have C III 4650 emission... See references in ApJ 711, L143.  E.g. HD 138403, very nice.  %Maybe we should look at some of them sometime...  Cheers!

\section{Spectral variability and period}
\label{sect_var}

\begin{figure}
\includegraphics[width=8cm]{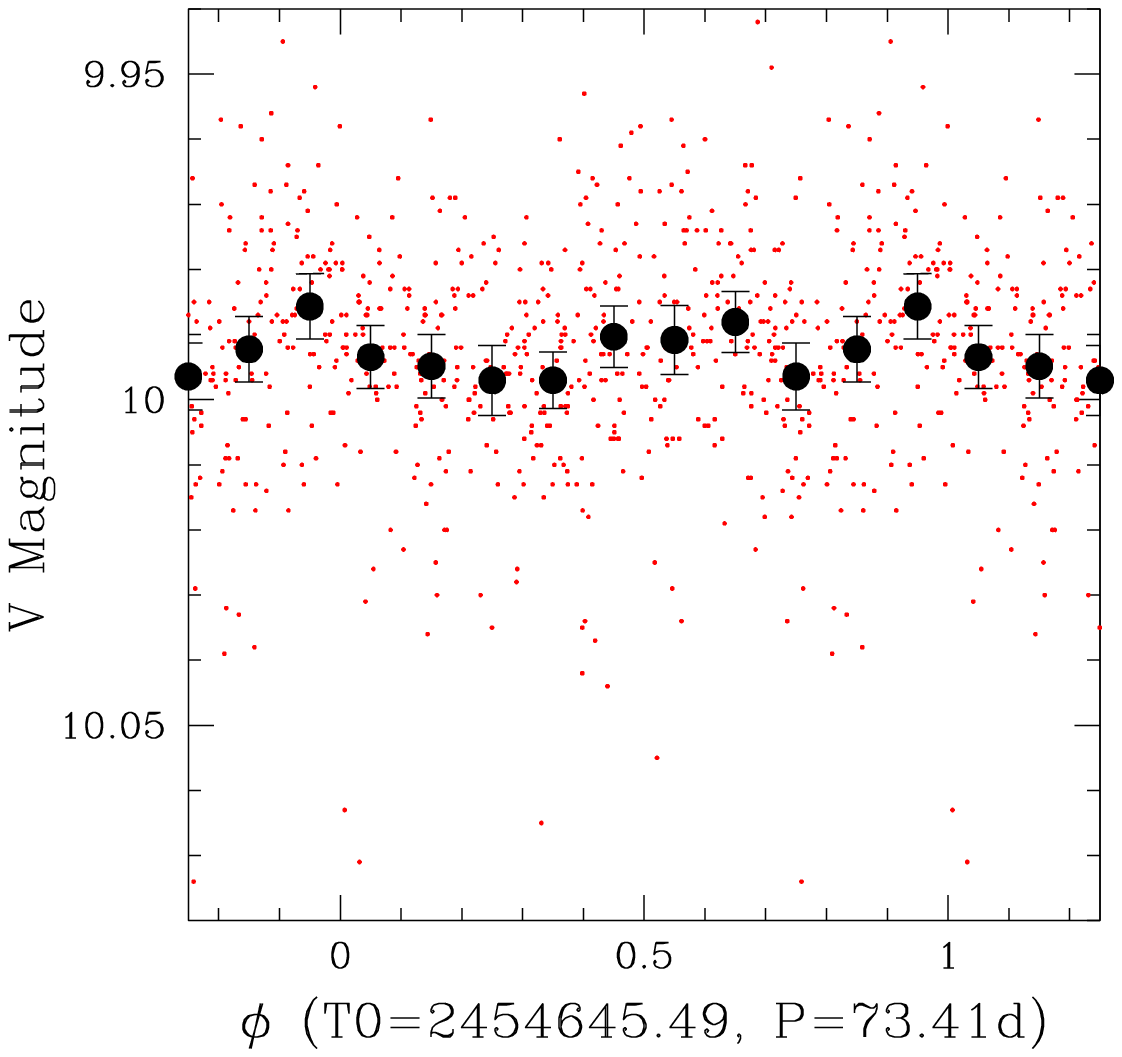}
\caption{Binned ASAS photometry phased with the spectroscopic ephemeris given by Eq. (1). Only a marginal variation is detected.}
\label{phot}
\end{figure}

\begin{figure}
\includegraphics[width=8cm]{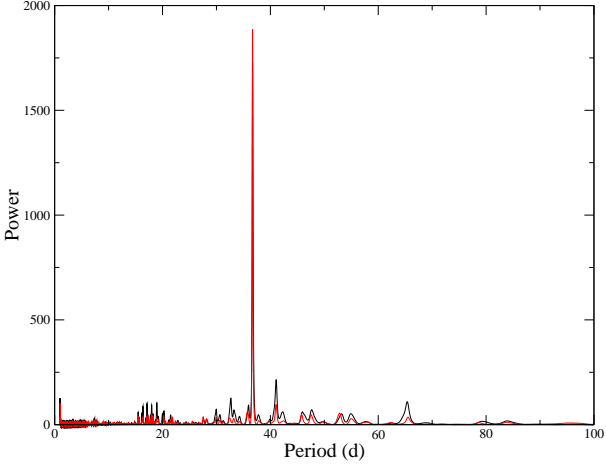}
\caption{Periodogram obtained from the H$\alpha$ EW (solid black) and He\,{\sc ii} $\lambda$4686 EW (solid red) measurements. A clear and unique signal at 36.7\,d is detected.}
\label{periodogram}
\end{figure}

Many lines in the spectrum of \cpd\ show significant variability. 

%Basic discussion of spectral variability (similar to HD\,148937) to begin with (Gregg, do you want me to write this part too?).\\

Following a similar analysis as applied by the MiMeS collaboration to other magnetic O-type stars \citep[e.g.][]{2012MNRAS.426.2208G, 2012MNRAS.419.2459W}, we first characterise the line variability using the equivalent width (EW). Before measuring the EW, each spectral line was locally re-normalised using the surrounding continuum, and the EWs were computed by numerically integrating over the line profile. The 1$\sigma$ uncertainties were calculated by propagating the individual pixel uncertainties in quadrature. Precomputed pixel error bars were only available for the ESPaDOnS and HARPS spectra, which required us to assign a single uncertainty to each pixel for the other spectra that was inferred from the RMS scatter of the Stokes $I$ flux in the continuum regions around each spectral line. 

A period search was performed on the EW measurements from all spectra using the Lomb-Scargle technique \citep{1992nrfa.book.....P} and an extension of this technique to higher harmonics \citep{1996ApJ...460L.107S}. { In this approach, a harmonic function with free parameters $a_1, a_2, a_0$ and $P$ (the latter being the period) corresponding to the form $a_1\sin(\omega t) + a_2\cos(\omega t) + a_0$, where $\omega = 2\pi/P$, is fit to the data. Adding the first harmonic leads to the introduction of additional terms of the form $a_3\sin(2\omega t) + a_4\cos(2\omega t)$. The resulting periodograms are defined as the power $\sum\limits_{i=1}^4 a_i^2$ versus period $P$. }

\begin{figure}
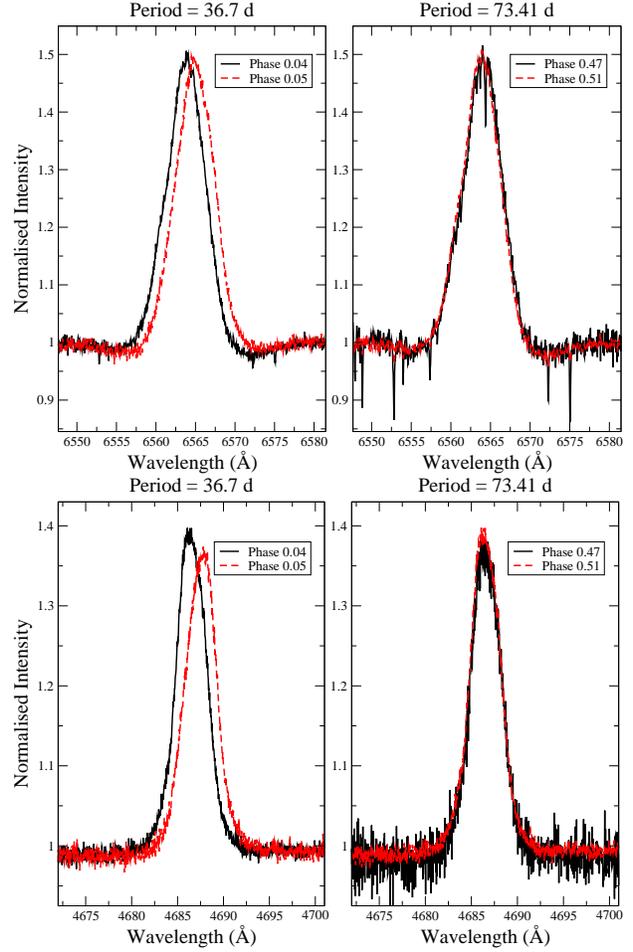

\includegraphics[width=3.2in]{halpha_comp.eps}
\includegraphics[width=3.2in]{he2_4686_comp.eps}
\caption{Comparison of H$\alpha$ and He\,{\sc ii} $\lambda$4686 profiles obtained at similar phases, but at different epochs. { {\em Left -}\ Assuming a period of 36.7\,d, profile shapes at similar phases do not agree. {\em Right -}\ Assuming a period of 73.4\,d, profile shapes at similar phases agree well.}  As discussed in the text, we identify the $P=73.41\pm0.05$ as the rotation period based on the better agreement of the profiles at all phases.}
\label{periodogram2}
\end{figure}

The periodograms from the strongly variable lines (such as H$\alpha$, H$\beta$, He\,{\sc i} $\lambda$5876 and He\,{\sc ii} $\lambda$4686) show significant power at $\sim$36.7\,d. When phased with this period, the EW variations show clear sinusoidal variations. However, the sinusoidal variations of the He\,{\sc i} $\lambda$5876 measurements show considerably more scatter than the other lines. Furthermore, a comparison of spectra obtained at different epochs, but at similar phases according to the $\sim$36.7\,d period, show { important differences. These include a} systematic velocity offset, a difference in peak emission flux, and opposite skew, of the emission line profiles. This effect is particularly striking for { H$\alpha$ and} He~{\sc ii} $\lambda 4686$, and is illustrated in Fig.~\ref{periodogram2}. { Lines in the spectrum of \cpd\ that are primarily in absorption are much more weakly variable; they are therefore poor probes of the wind variability reflected in the emission lines. Nevertheless, the behaviour of lines that are formed deeper in the photosphere (e.g. He~{\sc ii} $\lambda 4542$, C~{\sc iv} $\lambda\lambda 5801, 5811$) are indicative of a stable photospheric spectrum that is not plausibly responsible for the phenomenon discussed above.}

The detailed explanation is no different then the implementation used in fsrch:  

We therefore carried out a new period search including the contributions from the first harmonic - by including additional harmonics we change not only the relative power in each peak, but also the precise location of the peaks. 
The resulting periodograms showed two clear peaks, one consistent with the previous periodogram at $\sim$36.7\,d and a new peak at 73.41$\pm0.05$\,d, twice the previously identified period. When phased with this longer period, the EW variations { continue to} phase coherently. { However, this new phasing achieves a much better} agreement between the line profile shapes from spectra corresponding to similar phases, but obtained at different epochs. { We therefore conclude that a period of 74.41\,d provides a much better solution to the phase variability of \cpd, and we adopt this period} as the stellar rotation period within the context of a magnetically-confined wind and the oblique rotator model \citep{1950MNRAS.110..395S,1997ApJ...485L..29B,2011MNRAS.416.3160W}. Adopting maximum H$\alpha$ emission (minimum EW) as the reference date we derive the following ephemeris:
\begin{equation}
{\rm HJD}^{\rm max}_{\rm emis}  =  2454645.49(05) + 73.41(05)\cdot E,
\end{equation}
where the uncertainties (1$\sigma$ limits) in the last digits are indicated in brackets. { The uncertainties represent the formal 1$\sigma$ uncertainty computed from the $\chi^2$ statistic corresponding to second-order harmonic fits carried out on the EW variations for periods near 73\,d.}

\begin{figure*}
\includegraphics[width=2.3in]{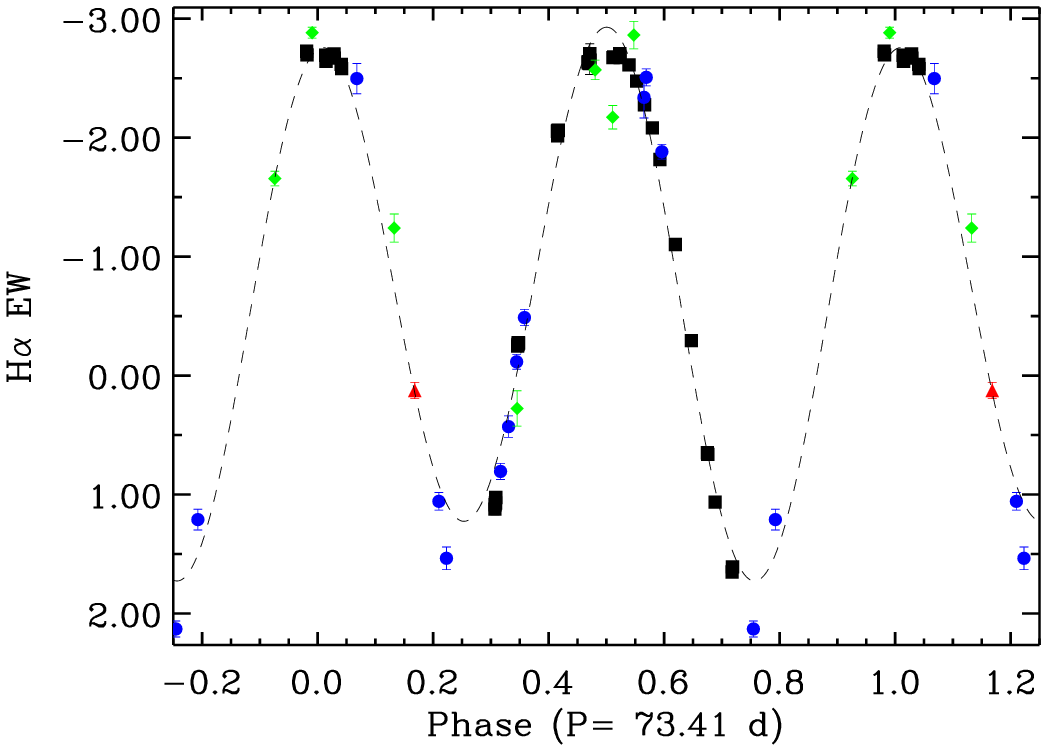}
\includegraphics[width=2.3in]{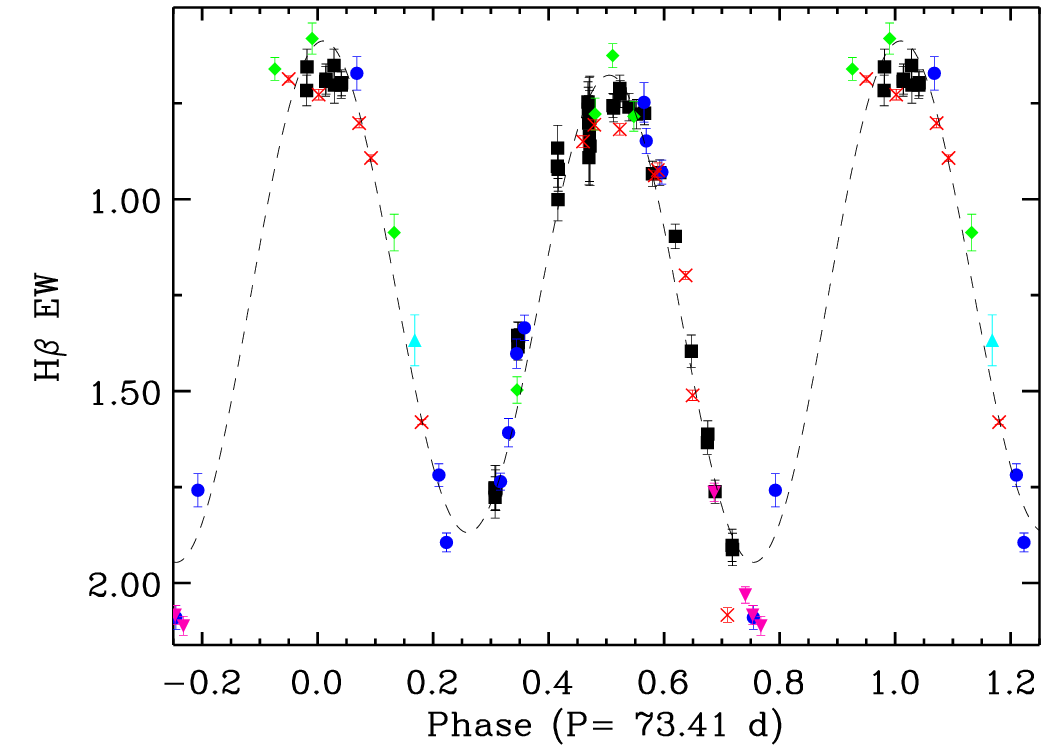}
\includegraphics[width=2.3in]{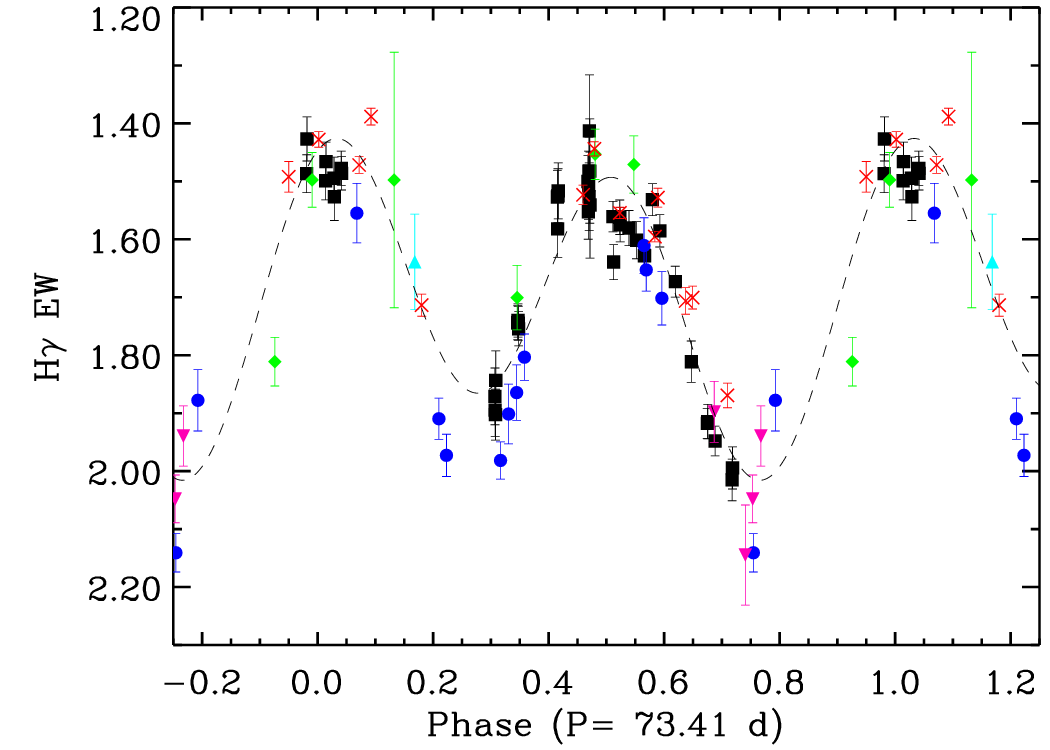}\\
\includegraphics[width=2.3in]{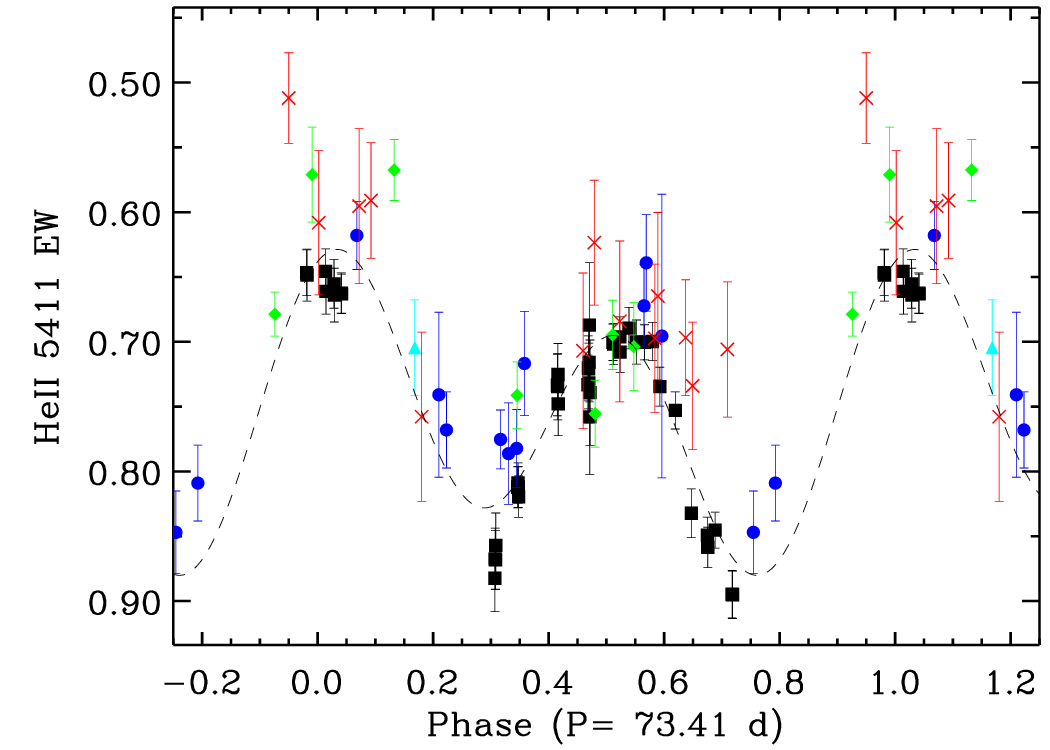}
\includegraphics[width=2.3in]{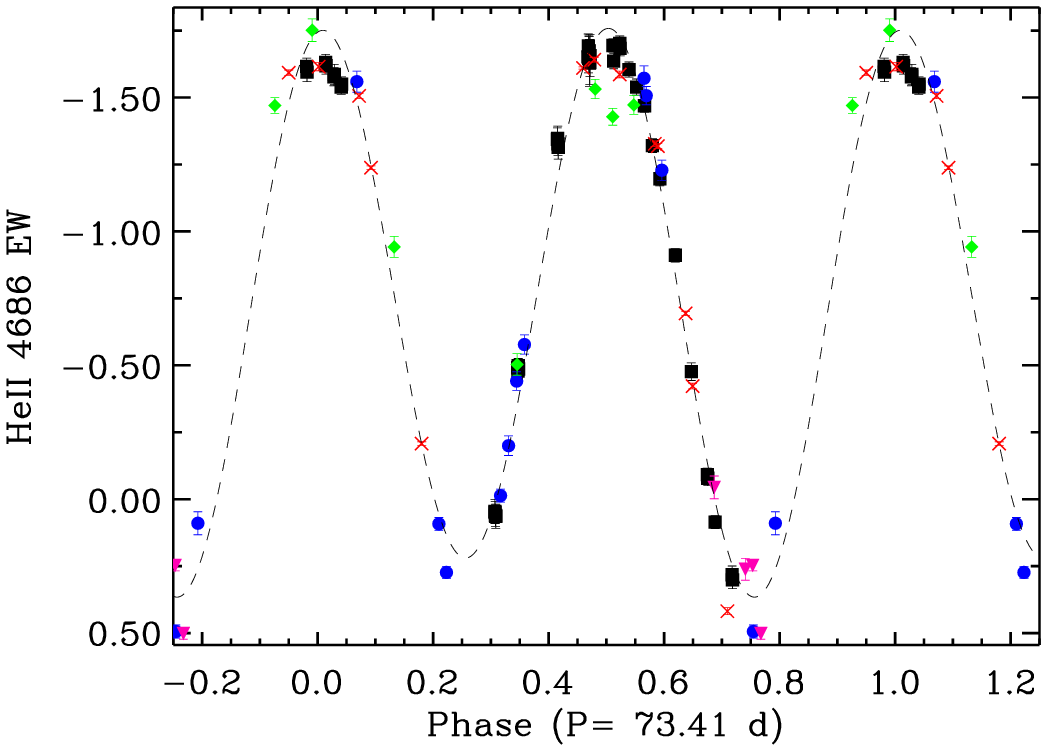}
\includegraphics[width=2.3in]{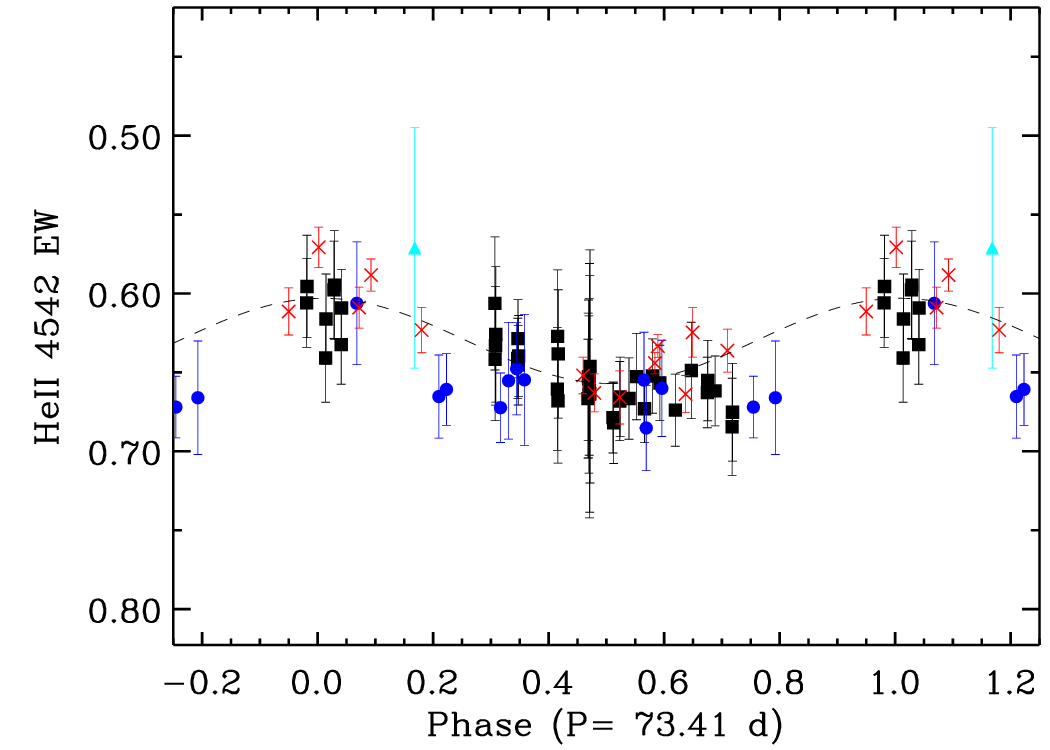}\\
\includegraphics[width=2.3in]{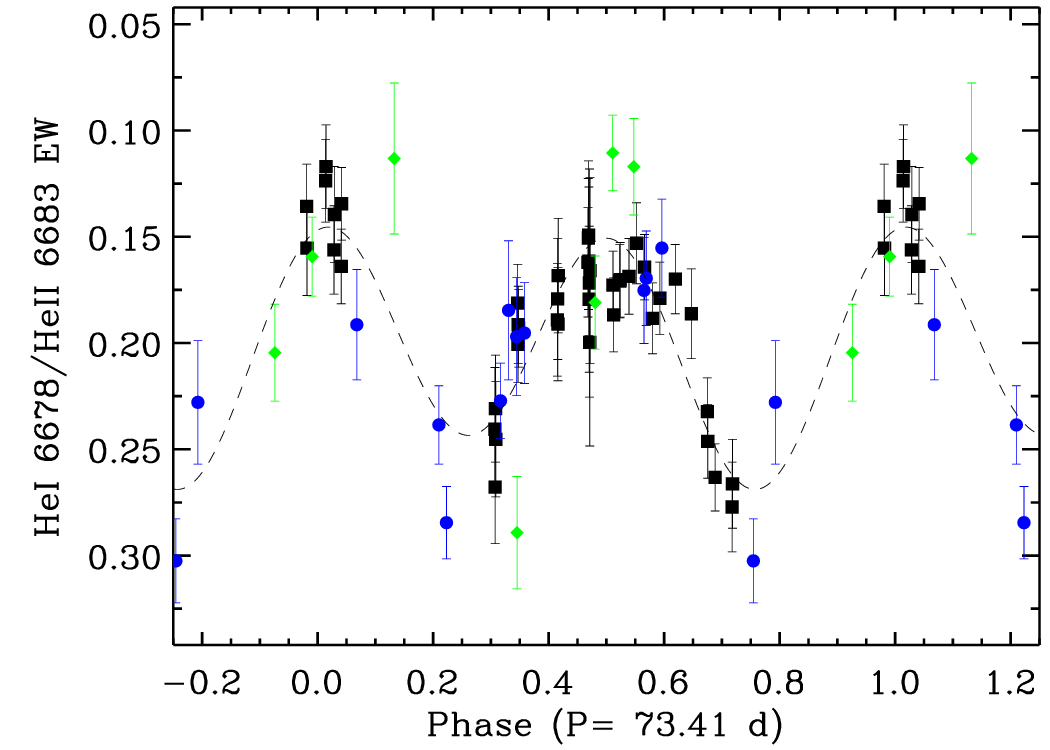}
\includegraphics[width=2.3in]{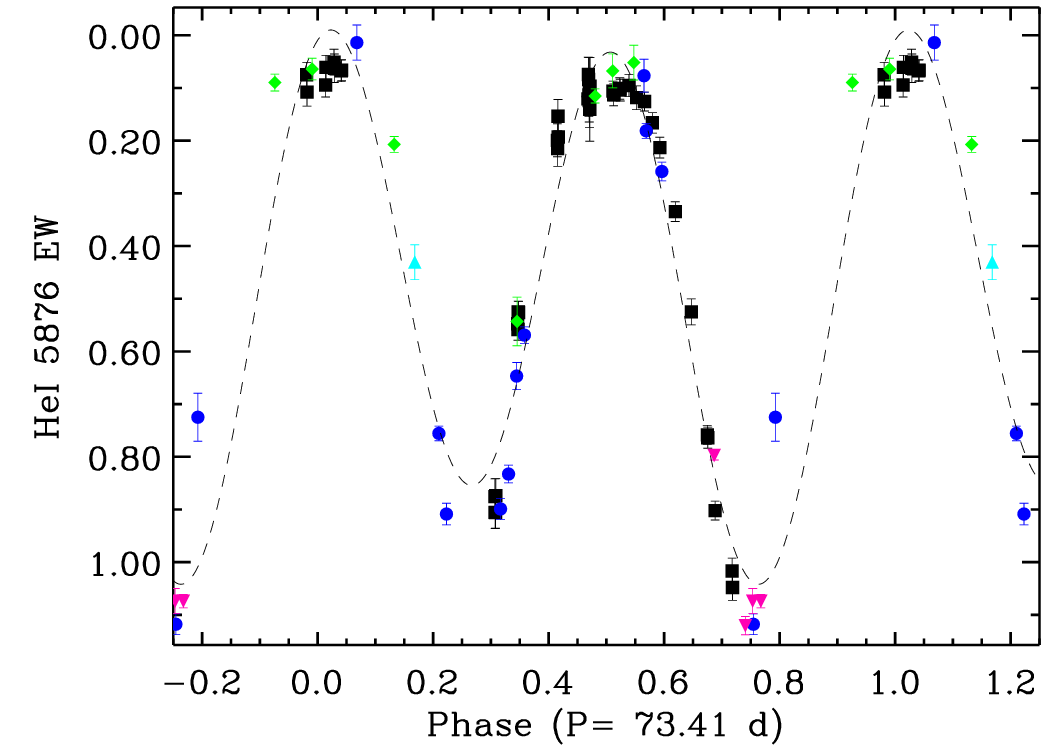}
\includegraphics[width=2.3in]{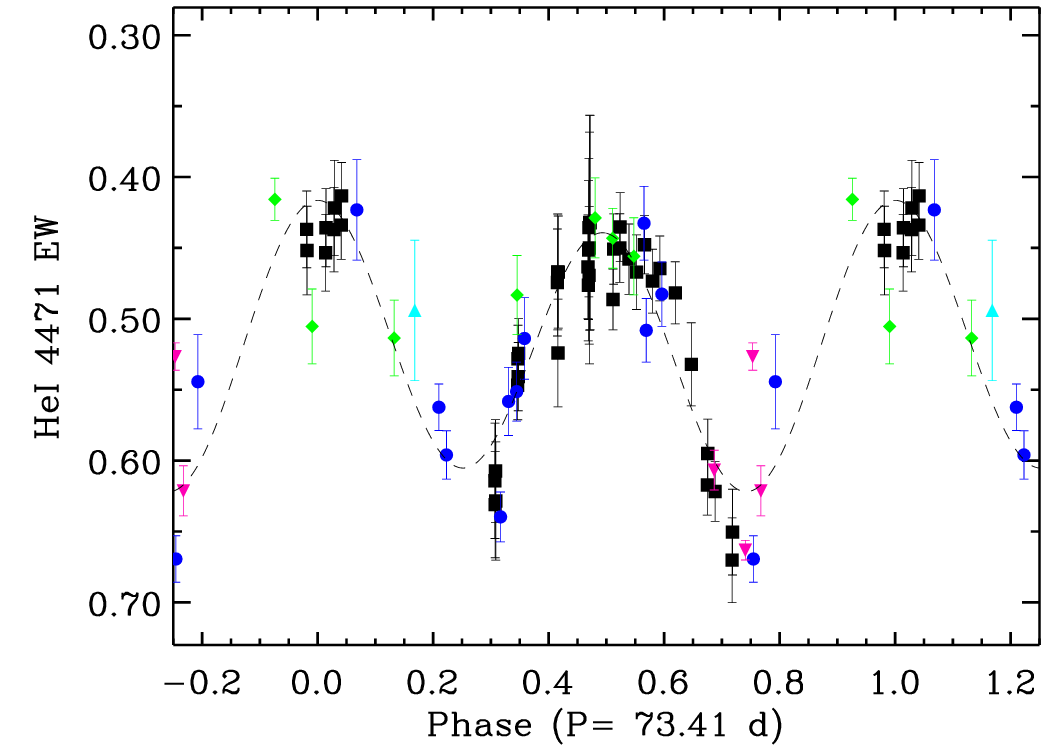}\\
\includegraphics[width=2.3in]{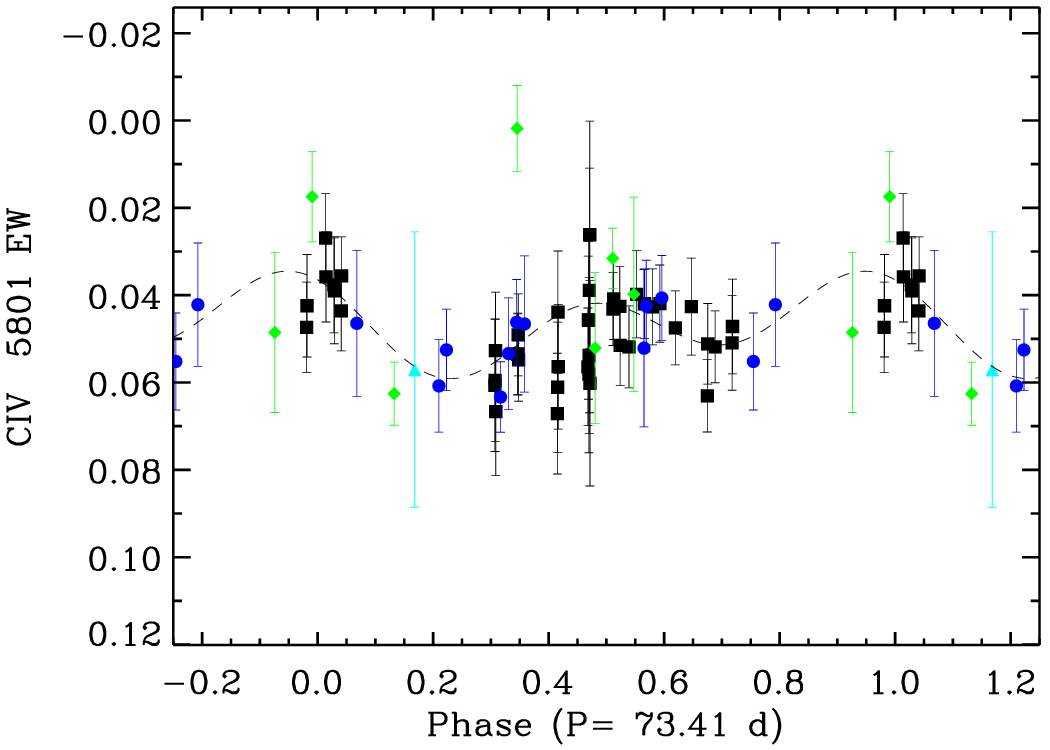}
\includegraphics[width=2.3in]{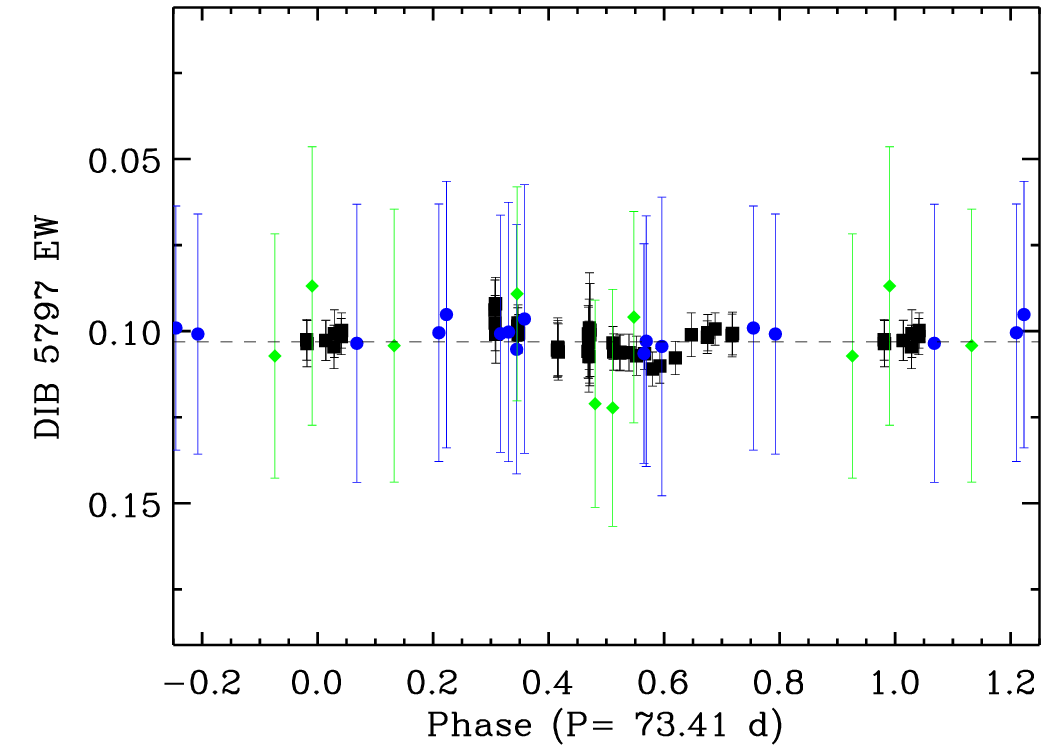}
\includegraphics[width=2.3in]{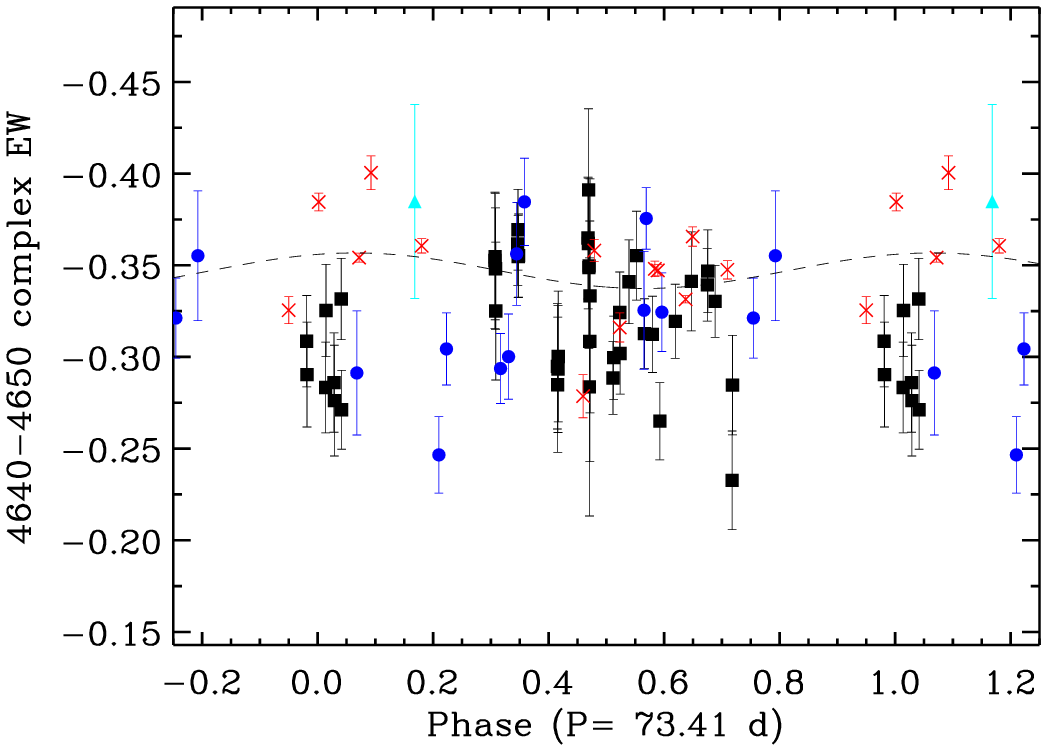}
\caption{Rotationally phased EW measurements for selected lines in the spectrum of \cpd. Least-squares sinusoidal fits are shown for those lines with statistically significant EW variability. In the bottom row we also show measurements of the $\lambda 5797$ DIB as an example of an intrinsically non variable feature. { In these figures, the ESPaDOnS measurements are black squares, the HARPS measurement is the cyan upwards facing triangle, the FEROS measurements are the blue circles, the LCO/echelle measurements are the green diamonds, the CAS measurements are the pink downward facing triangles, and the B\&C measures are red crosses.}}
%{ NOTES:} H$\alpha$: Offset added to LCO measurement of +0.3. H$\beta$: Offset added to CAS measurement of -0.15. H$\gamma$: Offset added to LCO dataset of 0.3. He\,{sc ii} $\lambda$4686: Offset of 0.05 added to CAS dataset. He\,{\sc i} $\lambda$5876: Offset of 0.1 added to CAS dataset.
\label{eqw_fig}
\end{figure*}

The phased EW measurements are illustrated in Fig.~\ref{eqw_fig}. { The measurements obtained from the various spectral datasets agree reasonably well.} Examination of Fig~\ref{eqw_fig} shows that most lines with significant variability exhibit double-wave variations. H$\alpha$ shows the most significant variability with a peak-to-peak EW variation of $\sim$5\ \AA. Maximum emission occurs at phase 0 for most lines, while another emission peak occurs one-half of a cycle later. The value of the EW measurements of H$\alpha$ and He\,{\sc ii} $\lambda$4686 at maximum emission are similar at both emission maxima, while a higher emission level occurs at phase 0.5 for some other lines (this is most evident in the EW curves of H$\beta$ and He\,{\sc i } $\lambda$5876), although there is considerable scatter at these phases. Unlike HD\,148937, which showed clear variability in C\,{\sc iii} $\lambda$4647 and C\,{\sc iv} $\lambda$5811 \citep{2012MNRAS.419.2459W}, \cpd\ shows no significant variability in these lines. We also include EWs measured from the DIB at 5797\,\AA\ to illustrate the lack of variability of a reference line that is not formed in the environment of the star.

Hipparcos and ASAS data are also available for this star, and we used these data to attempt to detect photometric variability and confirm the derived period. Only the best quality data were kept : for Hipparcos, this means keeping only data with flag 0; for ASAS data, two filterings were used - either keeping only data with grade A (and discarding four strongly discrepant points with $V>10.1$ mag) or keeping data only at 3 $\sigma$ from the mean (both filtering yielded the same results). Errors amount typically to 0.03 mag for $H_{\rm p}$ (Hipparcos), 0.2 mag for $B_{\rm T}$ and $V_{\rm T}$ (Tycho), and 0.04 mag for $V$ (ASAS). A $\chi^2$ test for constancy was performed on each dataset. Only $H_{\rm p}$ data were found to be significantly variable with significance level $SL<1$\%. However, a Fourier period search on those data reveals only white noise, without any significant peaks (or in particular a peak at the expected period of 73.41d). Furthermore, that period does not yield a significant coherent phased variation. Fourier period searches on the Tycho data yield similar conclusions. Period searches (Fourier \citep{2001MNRAS.327..435G}, PDM \citep{1978ApJ...224..953S}, and entropy \citep{1995ApJ...449..231C}) on the ASAS data result in the detection of marginal signals with periods of 63.69 $\pm$ 0.13d and 73.43$\pm$0.16d. The amplitude is very small, $\sim 0.0046$~mag at most. When phased with these periods, the photometry only shows marginal variations, even after binning the phased data (Fig.~\ref{phot}). We therefore conclude that any astrophysical variability of the photometric signal is below the typical uncertainty of a few tens of mmag, and would require high-precision photometry to be securely detected.

\begin{figure*}
\includegraphics[width=2.3in]{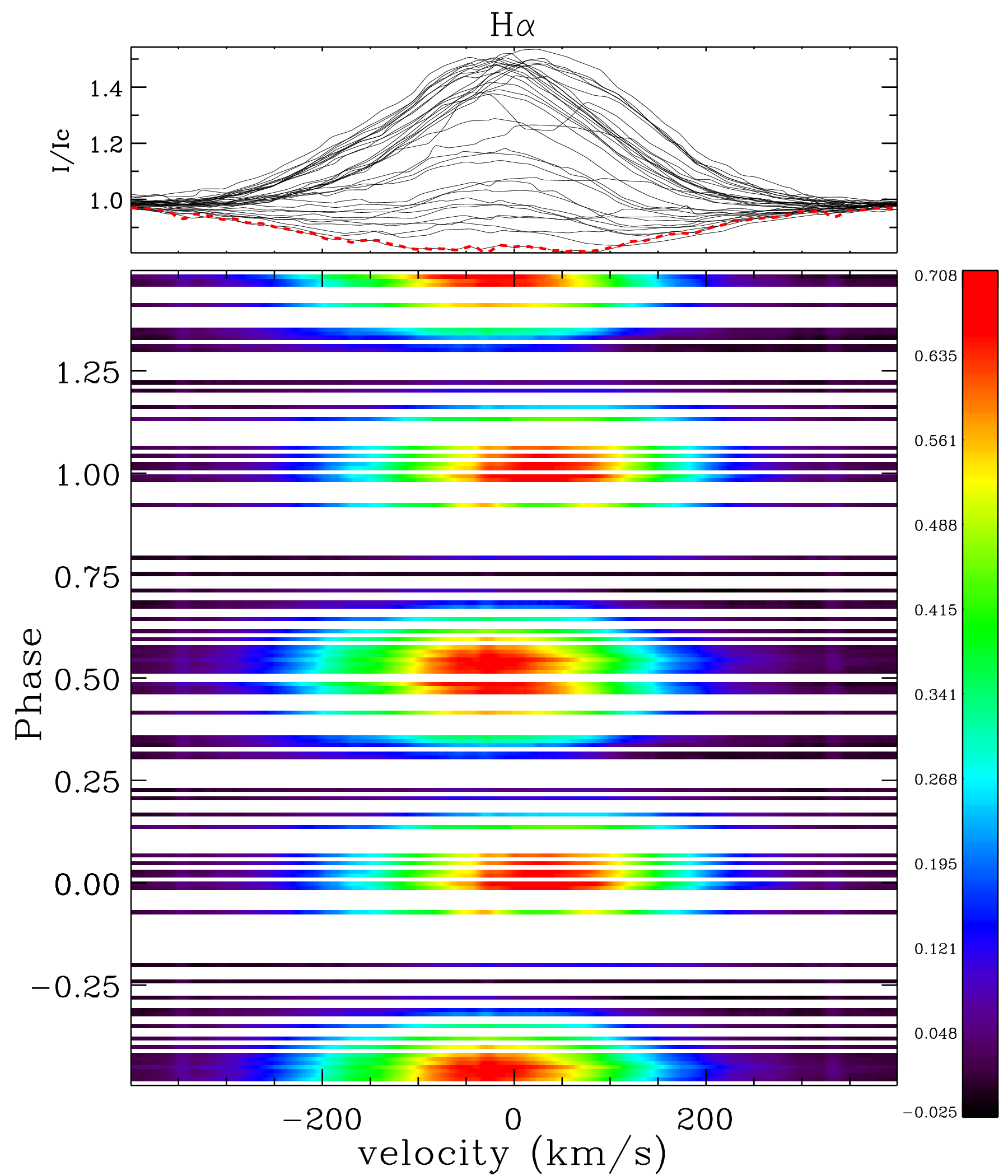}
\includegraphics[width=2.3in]{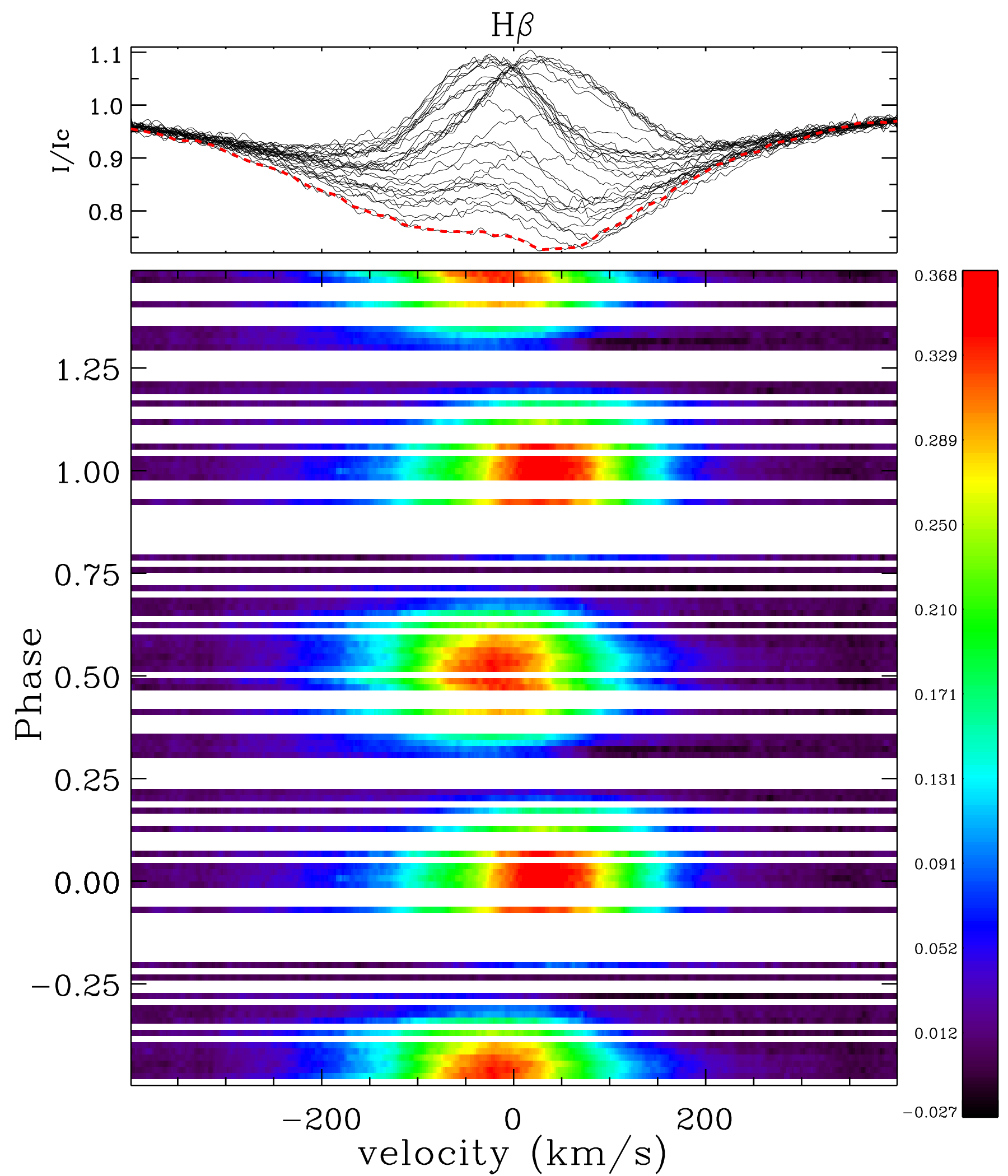}
\includegraphics[width=2.3in]{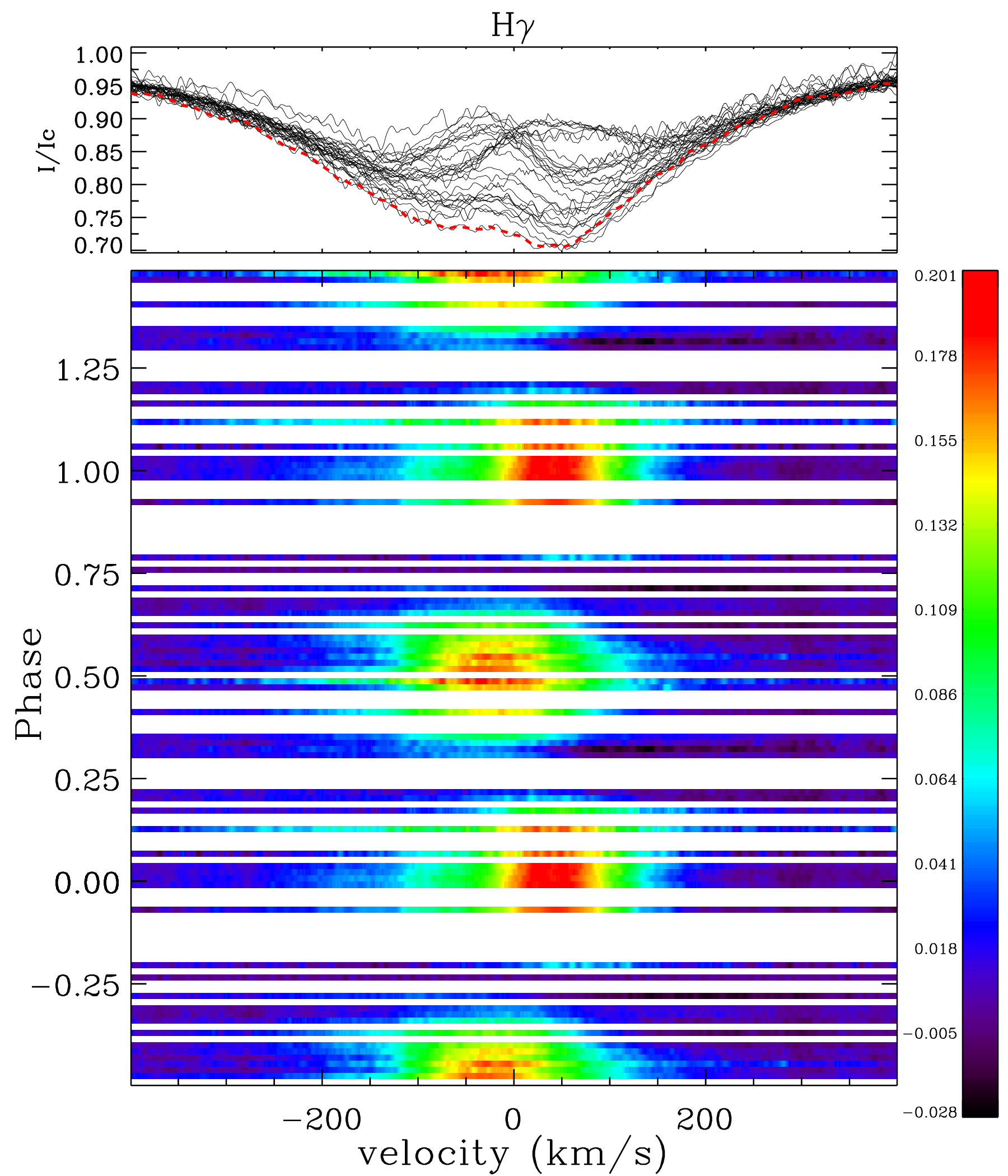}\\
\includegraphics[width=2.3in]{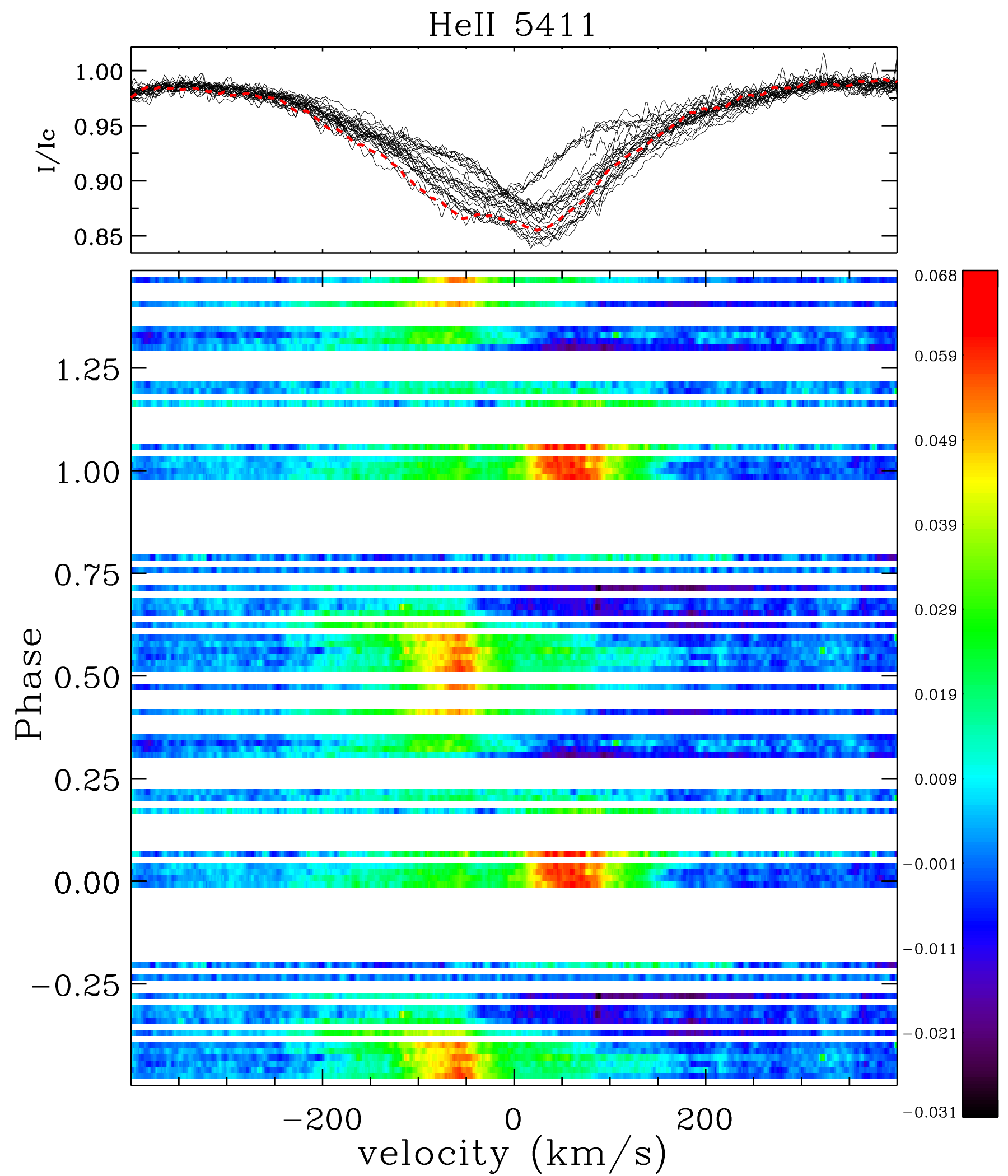}
\includegraphics[width=2.3in]{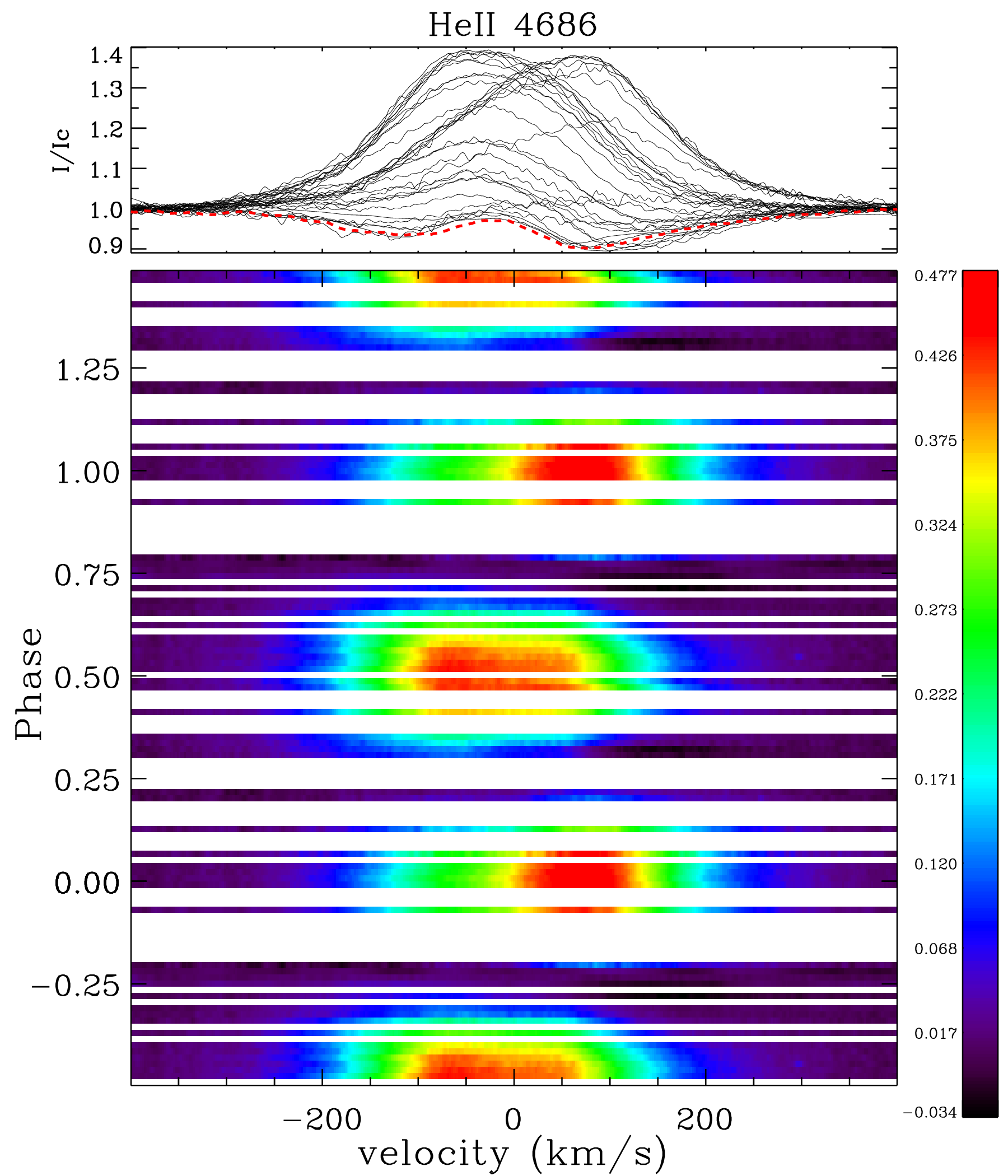}
\includegraphics[width=2.3in]{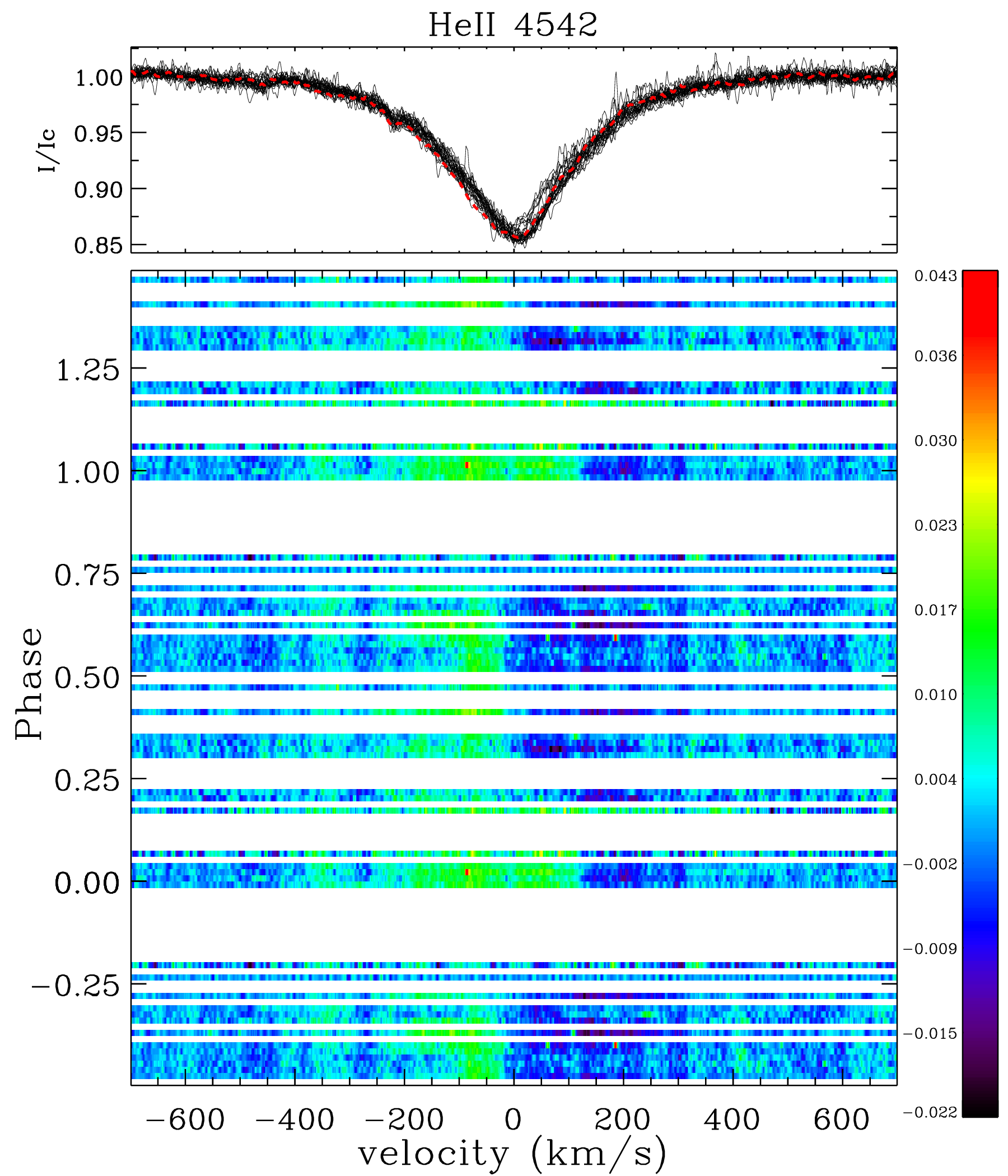}\\
\includegraphics[width=2.3in]{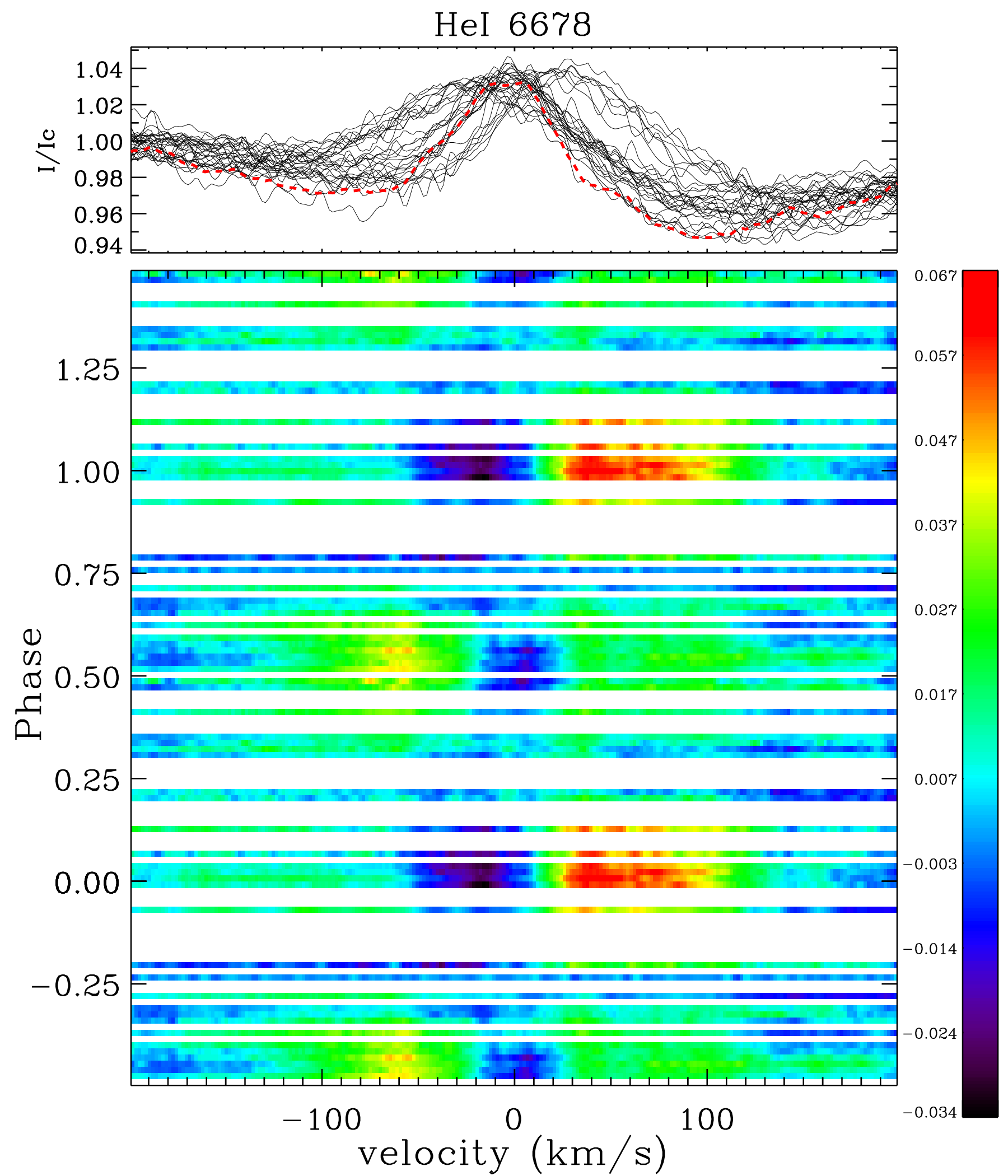}
\includegraphics[width=2.3in]{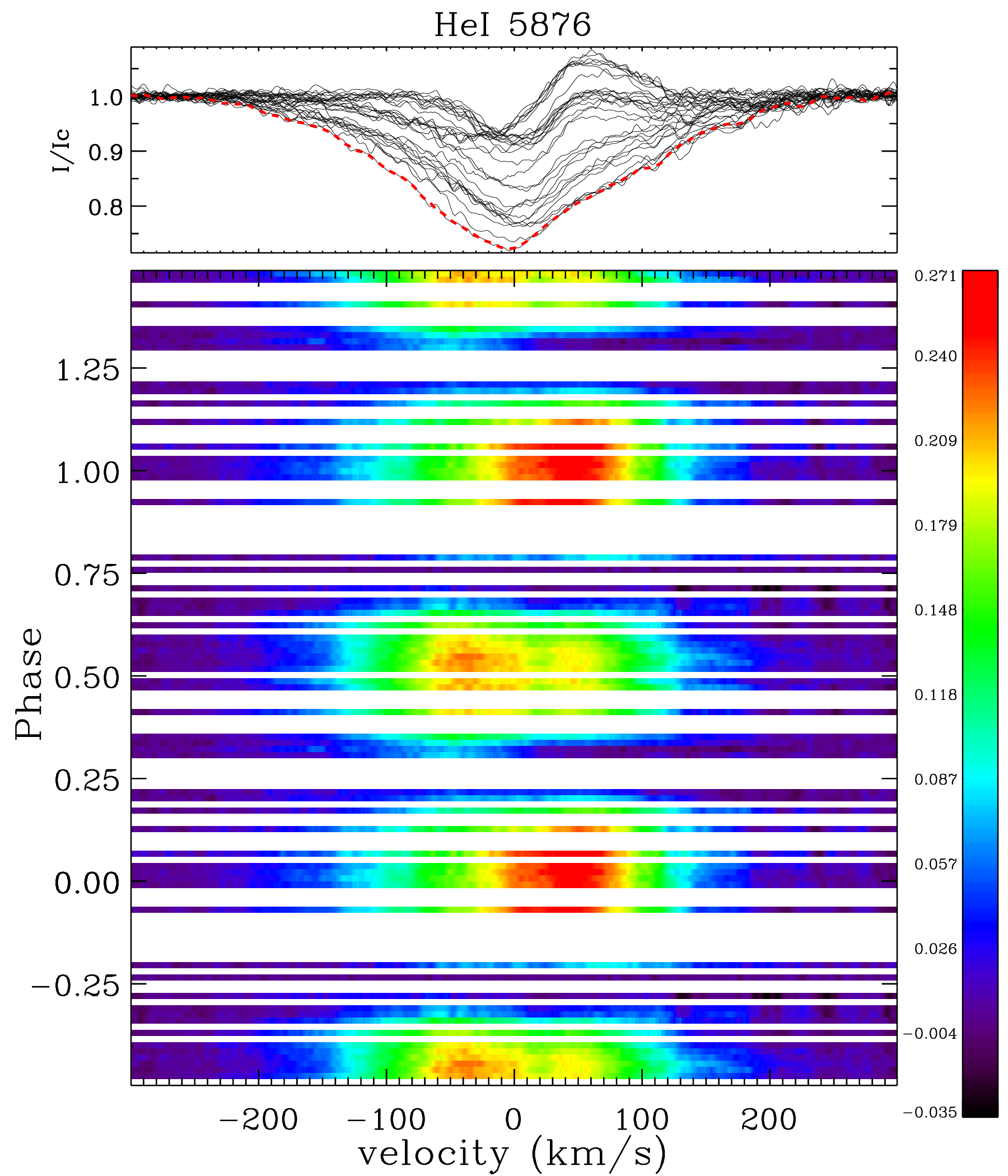}
\includegraphics[width=2.3in]{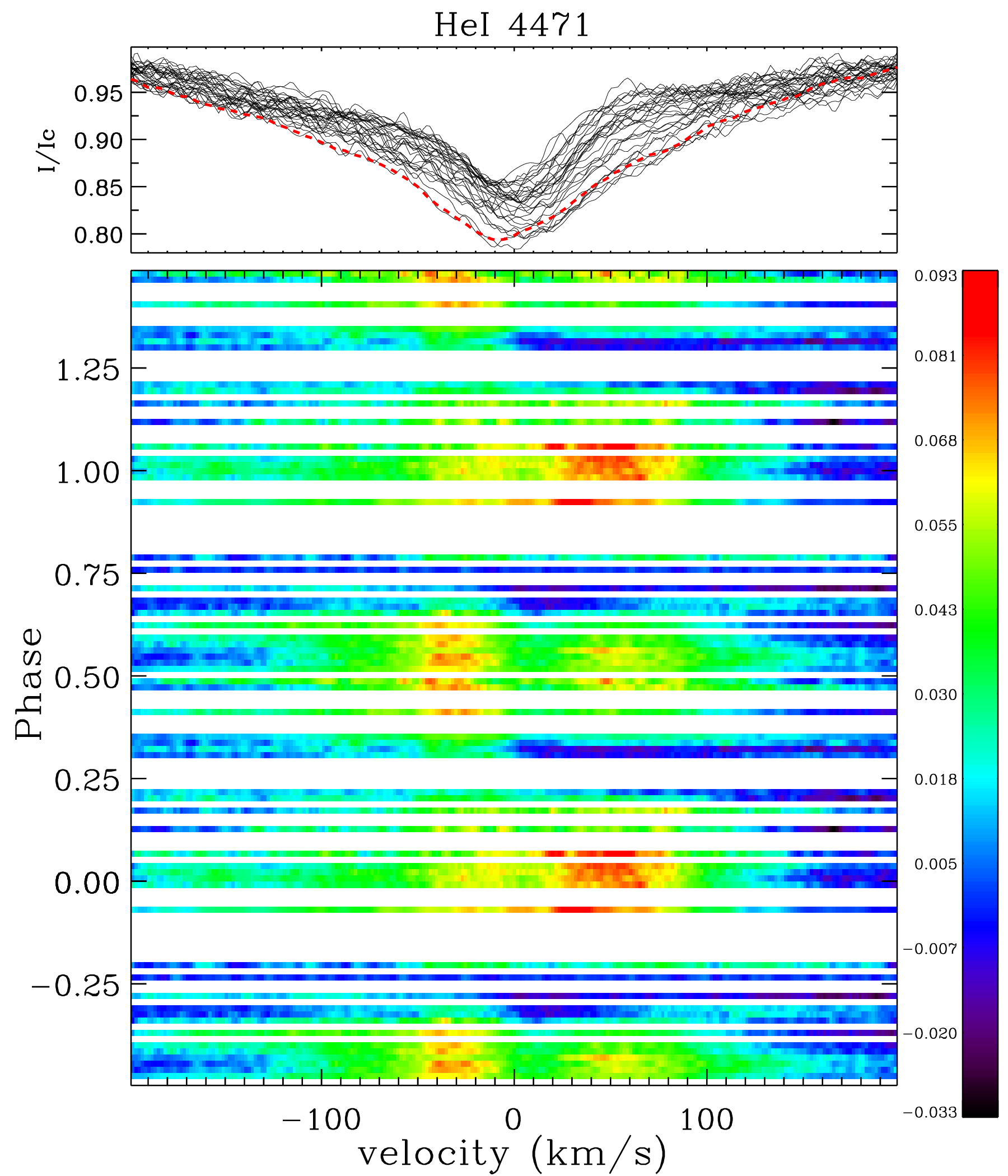}
\caption{Phased variations of selected spectral lines { in the high resolution spectra}, shown as dynamic spectra. Plotted is the difference between the observed profiles and the profile obtained of HJD 2454627.467  (dashed-red, top panel), which occurred at phase $\sim$0.75, corresponding to minimum observed emission in H$\alpha$.}
\label{dyn_fig}
\end{figure*}

%\begin{figure*}
%
%\includegraphics[width=2.3in]{c3_4647_dynamic.eps}
%\contcaption{}
%\end{figure*}

\subsection{Line profile variations}

\begin{figure*}
\begin{centering}
\includegraphics[width=14cm,angle=-90]{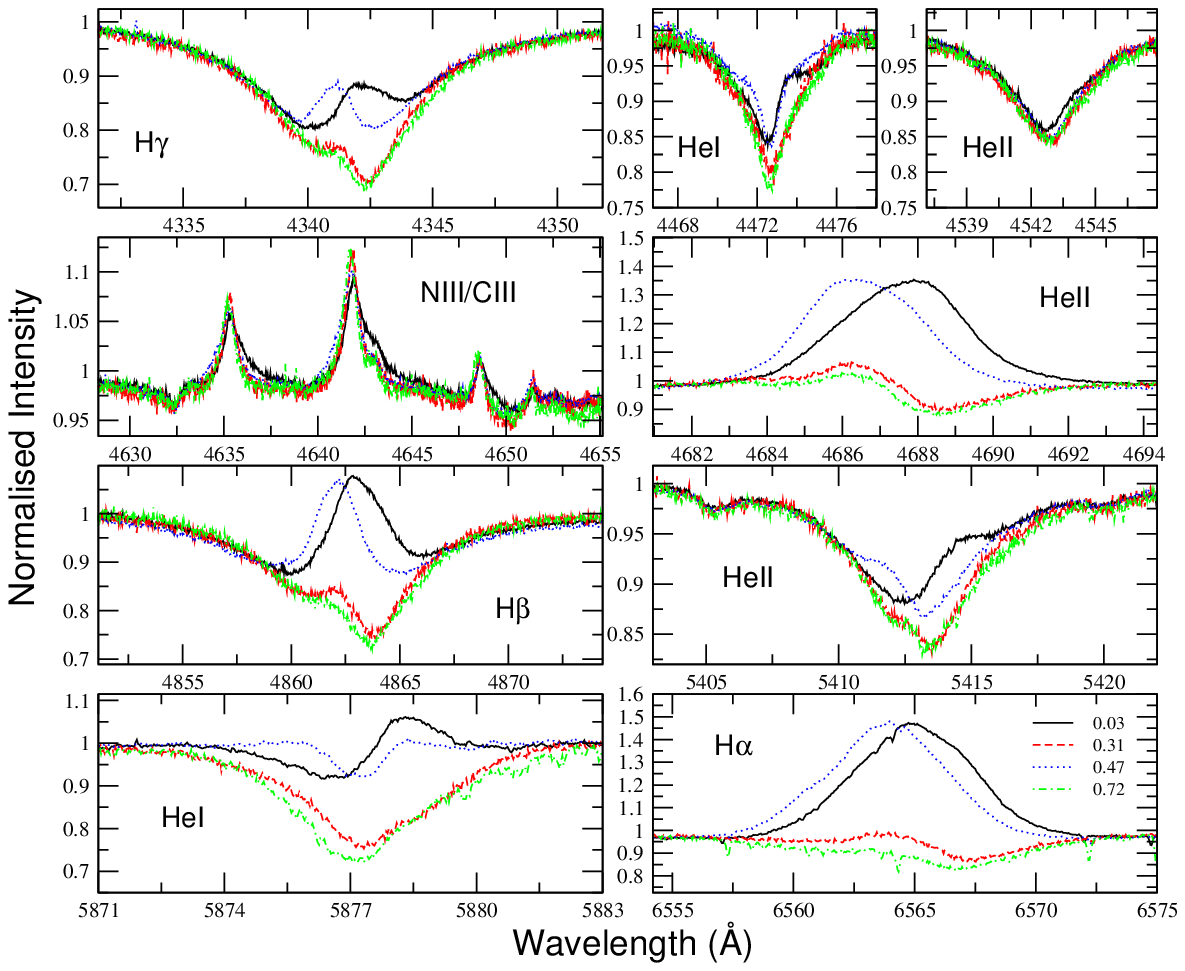}
\caption{\label{fig_overview}Selected line profiles in the spectrum of \cpd\ and their variability illustrated near the 4 principal phases according to the ephemeris derived in Sect.~\ref{sect_var}.}
\end{centering}
\end{figure*} 

\subsubsection{Line profiles of \cpd\ at high resolution}

As discussed above, it is now established that \cpd\ is the only Of?p star known so far that has double emission-line maxima, corresponding to a geometry that presents both magnetic poles during the rotational cycle.  The magnetic O stars HD 47129 \citep{2013EAS....64...67G} and HD~57682 \citep{2012MNRAS.426.2208G} do likewise. In the former case, the complex spectrum of this SB2 system precludes any identification of Of?p characteristics. In the latter case, its wind density is too low to produce the defining He~{\sc ii}, C~{\sc iii}, and N~{\sc iii} emission lines, while it does display the characteristic, variable Balmer emission profiles (first noted in a single H$\alpha$ observation by \citet{1980ApJS...44..535W}.  Thus, HD~57682 does provide points of comparison for \cpd.

An early indication that the double-wave period of $\sim 73$~d is the correct one was provided by the strikingly opposite skews of the He~{\sc ii}~$\lambda$4686 emission peaks at phases 0 and 0.5 (Fig.~\ref{fig_overview}).  In addition, there is a velocity offset between the two maximum phases.  Both of these effects are also seen in the Balmer emission lines to a lesser degree.  Significantly, both effects are likewise seen at H$\alpha$ in HD~57682.  These fine morphological details undoubtedly code significant information about the complex circumstellar structures producing them. 

The He~{\sc ii}\ $\lambda$4686 and Balmer profiles at the intermediate phases (i.e., at or near 0.25 and 0.75) are also noteworthy but display diversity between the two objects.  In \cpd, they all have weak emission near the blueward edges of absorption features at both intermediate phases.  As already noted, HD~57682 has no $\lambda$4686 emission, but remarkably, the H$\alpha$ profiles have the weak emission components shifted in {\it opposite} directions at the two intermediate phases. Again, these subtle similarities and differences are important physical clues that need to be modeled and understood. 

Turning to the systematic variations of He~{\sc ii}\ $\lambda 4686$ during the \cpd\ cycle, some further interesting effects are evident.  The weak emission toward the shortward sides of the intermediate-phase profiles gradually strengthens and shifts toward the line centre beginning about 0.3 cycle before each maximum (which again, have opposite skews and a small velocity difference). In contrast, however, the disappearance of the maximum emission lines is remarkably abrupt, occurring within about 0.1 cycle.  This behavior may be suggestive of a sharp occultation of the line-emitting region. 

%It is also noteworthy that the C~{\sc iii} emission, while always weak as discussed above, has its maximum strength relative to the N~{\sc iii}, approaching that of $\lambda 4634$, at the odd phases of 0.1 and 0.6. { Nolan, what are the origin of these phases? What is the evidence that these lines show maximum emission at phases other than 0.0 and 0.5? Please update theses phases are required.}

The behaviour of the dilution-sensitive line He~{\sc i}\ $\lambda 5876$ is entirely distinct from those of the features just described, which are either similar or opposite between phases 0.5 apart.  That of $\lambda$5876 is egregiously asymmetrical (Fig.~\ref{fig_overview}): at phase 0.5 it displays a weak absorption flanked by equal emission wings, whereas at phase 0.0 it has a relatively weak but well marked P Cygni profile.  At phases 0.25 and 0.75 it is a strong, symmetrical absorption line. 

On the other hand, He~{\sc ii} $\lambda$5411 is a moderate absorption line with strong wings at all phases, which at first glance appears to present velocity shifts {opposite} to those of the $\lambda$4686 and Balmer emission lines at the two maxima.  However, on closer inspection the effect is seen to be due to wing emissions in the same sense as those of the other lines, although possibly with a larger amplitude.  In contrast, C~{\sc iv} $\lambda\lambda$5801, 5812 are weak absorption features with no velocity shifts, albeit with a weak P Cygni tendency at phase 0.0, i.e., in the sense of $\lambda$5876.  No doubt this apparent chaos will be transformed into valuable diagnostics as our understanding of the intricate phenomenology advances. 

Finally, the relative behaviours of He~{\sc i} $\lambda$4471 and He~{\sc ii} $\lambda$4542 (Fig.~\ref{fig_overview}) are of considerable interest, since their ratio is the primary spectral-type criterion, although classification is a heuristic exercise here because of the effects of variable circumstellar emission on the line strengths; the minimum spectrum is expected to be more closely related to the actual stellar parameters.  Indeed, both lines are seen to be affected by wing emission at both maxima, albeit the He~{\sc i} line much more strongly.  In fact, the $\lambda 4542$ line has the unique characteristic that its profile at phase $\sim 0.5$ is very similar to that exhibited at the quadrature phases 0.25 and 0.75. Hence it appears that this line is only affected by one of the emission maxima. 

It should be noted that the He~{\sc i} effect is weaker than in other Of?p spectra, whereas the He~{\sc ii} one is unprecedented, weaker than but analogous to that in $\lambda$5411 described above. In terms of the line depths, the maximum spectral type is O7, but the He~{\sc ii} line has a larger equivalent width so that measurements would yield O6-O6.5.  On the other hand, the minimum spectral type is a well defined O8, which again may be presumed to correspond to the stellar photosphere.

\subsubsection{Dynamic spectra}

The phased line profile variations are displayed as dynamic spectra in Fig.~\ref{dyn_fig}. These images were constructed by subtracting the profile that presented the least overall emission in H$\alpha$ (corresponding to phase $\sim$0.75 - obtained on HJD 2454627.467). The colour scheme was chosen to maximise the dynamic range of the variability to highlight changes relative to the minimum profile.

%(nc0806ESO18751)

The dynamic spectra of the most strongly variable lines (such as the Balmer lines (H$\alpha$, H$\beta$, H$\gamma$), He\,{\sc i} $\lambda$5876 and He\,{\sc ii} $\lambda$4686) exhibit very similar characteristics. In each case the profiles indicate the presence of structures that vary in intensity, from absorption to strong emission, twice per cycle. The two emission features reach similar peak intensities (as already inferred from their EW variations). The emission features of the Balmer lines are slightly asymmetric in velocity about their central peaks and are found to be broad relative to the width of the spectral line (the emission features in H$\alpha$ have a FWHM of $\sim$280\,km\,s$^{-1}$). The emission feature that reaches maximum emission at phase 0 appears to be offset from the mean velocity by about -30\,km\,s$^{-1}$; the emission feature at phase 0.5 is offset by +30\,km\,s$^{-1}$. 

While the line profile variability in He\,{\sc i} $\lambda$5876 and He\,{\sc ii} $\lambda$4686 appears similar to the Balmer lines, there are some outstanding differences. The two emission features appear considerably more asymmetric. In fact, the emission feature occurring at phase 0.5 appears to be a blend of two distinct emission peaks in both lines. The higher intensity peak is centred around -60\,km\,s$^{-1}$ for the He\,{\sc ii} line (or $-35$\,km\,s$^{-1}$ for the He\,{\sc i} line), while the less intense peak is centred about 30\,km\,s$^{-1}$ for the He\,{\sc ii} line (or 50\,km\,s$^{-1}$ for the He\,{\sc i} line). While the central velocities of these features differ between these two lines, their separation is similar. The emission features that appear at phases 0 and 0.5 reach similar peak intensity for the Balmer lines, whereas the intensity of the emission peak at phase 0 is about 10\% stronger than the peak occurring at phase 0.5 for the He~{\sc ii} line and 20\% stronger for the He~{\sc i} line.

%Unlike the Balmer lines where the relative intensity of the emission features were nearly identical, the intensity of the strongest emission peak at phase 0 is about 10 percent stronger than the peak occurring at phase 0.5 for the He\,{\sc ii} line, but is 20 percent stronger for the He\,{\sc i} line.

The lines that display weaker variability exhibit a character of variability that is similar to those described for the previous lines. The He\,{\sc i} $\lambda$6678 line shows weaker emission than the previously discussed spectral lines, but there is still evidence for an emission feature occurring twice per cycle. This feature occurs at a velocity relative to the mean velocity that is similar to the other He lines. Furthermore, the relative intensity of the redward emission feature appears considerably stronger than the blueward feature in the He\,{\sc i} lines that display less variability (the maximum emission intensity is about 60 percent stronger than the blueward emission feature of the $\lambda$6678 line). Some of the weakly-variable lines also show evidence of enhanced absorption (relative to the minimum profile) in a narrow region around the line core. The enhanced absorption reaches a maximum relative absorption in the core at about phase 0.5 (which corresponds to the phase of maximum emission at the blue edge of this line). 

%Lastly, while the C\,{\sc iv} $\lambda\lambda$5801, 5811 lines show evidence for similar emission features as previously discussed, we note that the blueward emission feature occurring at phase 0.5 appears to reach a higher maximum emission than the feature that occurs at phase 0, which is opposite behaviour to what is observed in the other lines.

The He~{\sc ii}\ $\lambda 4542$ line exhibits only weak variability. However, it stands out from essentially all other lines in the spectrum of \cpd\ due to its apparent single-wave variation, a phenomenon reflected in the line profiles examined in Sect. 4.1.1. In contrast to the double-wave variation of the other lines illustrated in Figs.~\ref{eqw_fig} and \ref{dyn_fig}, which show EW maxima at phases 0.0 and 0.5, the $\lambda 4542$ line appears to exhibit an EW maximum at phase 0.0, but a minimum at phase 0.5.

\begin{figure*}
\begin{centering}
\subfloat[][Phase 0.5]{\includegraphics[width=8cm]{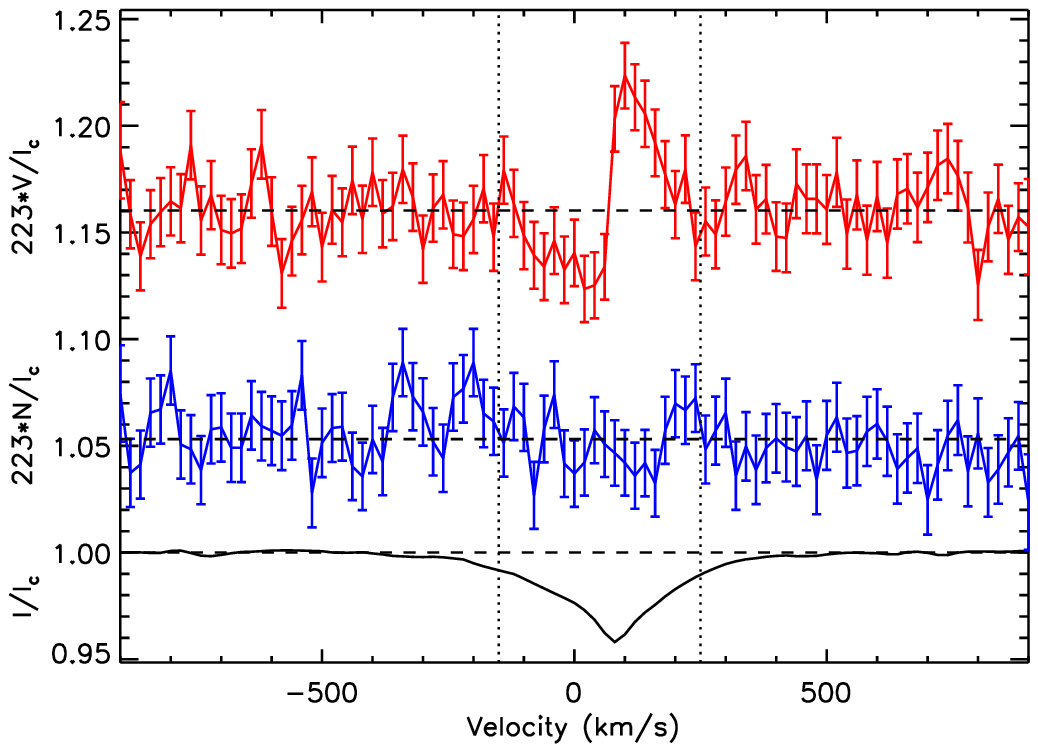}}\subfloat[][Phase 0.0]{\includegraphics[width=8cm]{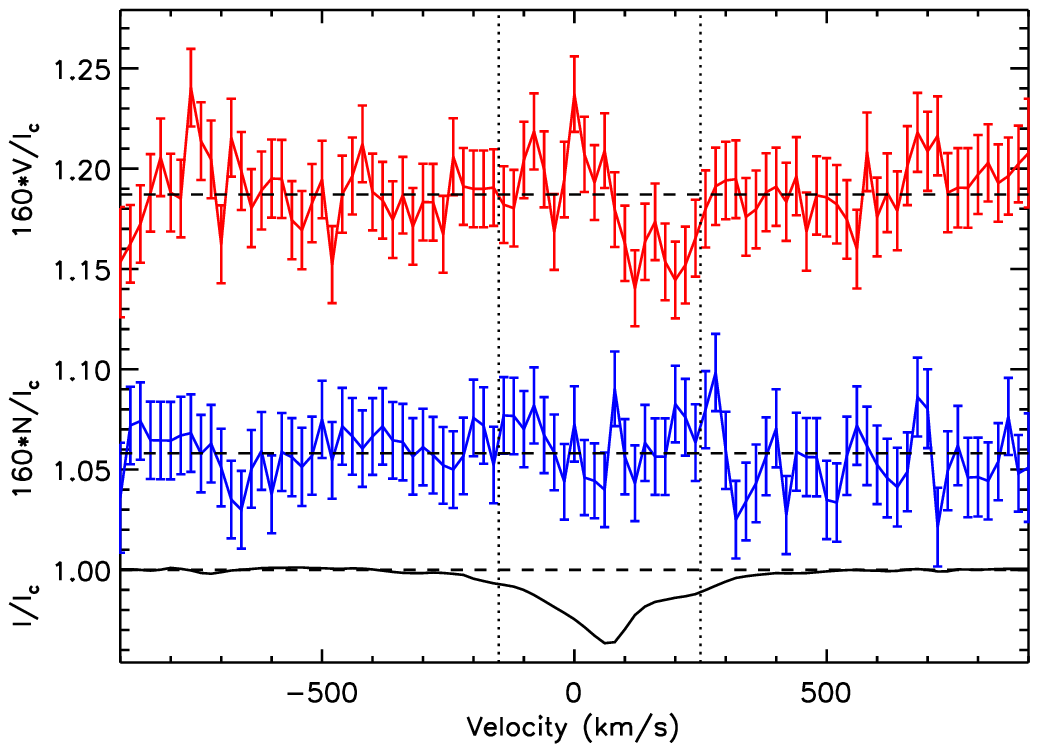}}
%\vspace{3in}
\caption{\label{lsdprofs}{\em Left:}\ LSD profile at phase 0.5, Stokes $V$: Definite detection (FAP=9e-7\%), $\bz=-290\pm 95$~G. Diagnostic null $N$: ND (FAP=72\%), $\nz=-2\pm 95$~G. {\em Right:}\ LSD profile at phase 0.0, Stokes $V$: No detection (FAP=2\%), $\bz=+335\pm 200$~G. Diagnostic null $N$: ND (FAP=91\%), $\nz=-71\pm 199$~G.}
\end{centering}
\end{figure*} 

\section{Diagnosis of the magnetic field}

%Since the field is so marginally detected, should probably bin results as early as possible.

%\subsection{Least Squares Deconvolution of the ESPaDOnS spectra}

Least-Squares Deconvolution \citep[LSD, ][]{1997MNRAS.291..658D} was applied to all CFHT observations using the LSD code of \citet{2010A&A...524A...5K}. In their detection of the magnetic field of HD 191612, \citet{2006MNRAS.365L...6D} developed and applied an LSD line mask containing 12 lines. Similar masks were successfully employed by \citet{2011MNRAS.416.3160W,2012MNRAS.419.2459W} in their analyses of HD 191612 and HD 148937. { Given the similarity between the spectra of \cpd\ and HD 191612, we began with this line list and adjusted the predicted line depths to best match the depths observed in the spectrum of \cpd\ at phases $\sim 0.25$ and $0.75$, when the emission is lowest. This involved adjustment of the line depths by typically $\sim 20\%$ relative to their depths in the spectrum of HD 191612.} We then used this line list to extract mean circular polarization (LSD Stokes $V$), mean polarization check (LSD $N$) and mean unpolarized (LSD Stokes $I$) profiles from all collected spectra. All LSD profiles were produced on an 1800~km\,s$^{-1}$ spectral grid with a velocity bin of 20 km\,s$^{-1}$, using a regularization parameter of 0.2 (for more information regarding LSD regularisation, see \citet{2010A&A...524A...5K}.

% Appropriate folder is ilsd_hd191612mask_maxabs20

Using the $\chi^2$ signal detection criteria described by \citet{1997MNRAS.291..658D}, we evaluated the significance of the signal in both the Stokes $V$ and $N$ LSD profiles in the velocity range [-150, 250]~km\,s$^{-1}$, consistent with the observed span of the Stokes $I$ profile. No significant signal was detected in any of the individual $V$ (or $N$) profiles. We also computed the longitudinal magnetic field from each profile set using Eq. (1) of \citet{2000MNRAS.313..851W}. To improve our sensitivity, we coadded the LSD profiles of spectra acquired within $\pm 2$ nights (resulting in 9 averages of 2-8 spectra; see Table~\ref{tab:specpol}). Given the observed variability period (Sect. 5), $\pm 2$ nights corresponds to approximately 0.05 cycles - a timespan during which variability should be limited; indeed, this was verified empirically. From these profiles we obtain one marginal detection of signal (false alarm probability ${\rm FAP}<10^{-3}$) in the $V$ profiles (for the coadded profile corresponding to spectral IDs 1604363-1604387), and best longitudinal field error bars of $\sim 170$~G. The most significant measurement of the longitudinal field from the coadded profiles is $-362\pm 186$~G (2.0$\sigma$). The individual and coadded ESPaDOnS spectra and the corresponding longitudinal fields and detection probabilities are indicated in Table~\ref{tab:specpol}. { The longitudinal field measured from the HARPSpol spectrum was $280\pm 460$~G.}

{ We conclude that we fail to detect a magnetic field in individual and co-added Stokes $V$ spectra of \cpd. To proceed further, we note the similarity of the spectrum of \cpd\ to HD 191612 (and in particular its Of?p classification), and its periodic variability, combined with the reported detection of a strong magnetic field by \citet{2011A&A...528A.151H,2012IBVS.6019....1H}, strongly suggest that \cpd\ is an oblique magnetic rotator. This has been convincingly demonstrated for HD 191612 \citep{2011MNRAS.416.3160W}, and is consistent with the behaviour of other magnetic O-type stars \citep[e.g.][]{2012MNRAS.425.1278W,2012MNRAS.426.2208G,2012MNRAS.419.2459W}.}

{ We therefore} proceeded to bin the LSD profiles according to rotational phase as computed via Eq. (1) in order to increase the signal-to-noise ratio. We note that this phase binning implicitly assumes an oblique magnetic rotator, i.e. that the magnetic field variation proceeds according to the same period as the spectra (and other variations). We have binned the coadded LSD profiles from phases 0.307-0.619 and 0.981-0.042. From these profiles (illustrated in Fig.~\ref{lsdprofs}) we obtain a definite detection of signal in the Stokes $V$ profile (at phase $\sim 0.5$) and no detection (although with twice poorer SNR) at phase $\sim 0.0$, with no signal detected in the null profiles. Notwithstanding the lack of formal detection at phase 0, the Stokes $V$ LSD profile appears to exhibit a weak signature with polarity opposite to that at phase 0.5. The phase 0.0 and 0.5 LSD profiles yield longitudinal fields of $\bz=+335\pm 200$~G and $\bz=-290\pm 95$~G respectively, and corresponding null profile fields of $\nz=-71\pm 199$~G and $\nz=-2\pm 95$~G. We have also searched the average spectrum at phase $\sim 0.5$ for Zeeman signatures in individual line profiles. Weak signatures, compatible with the LSD Stokes $V$ profiles, are visible in the He~{\sc i}~$\lambda 5876$ and C~{\sc iv}~$\lambda 5801$ lines. { We also note that coadding LSD profiles phased using one-half the adopted period (i.e. 36.7\,d) yields no detection of the magnetic field.}

From this analysis, { which implicitly assumes that \cpd\ is an oblique magnetic rotator}, we conclude that an organized magnetic field is detected in the photosphere of \cpd, with a longitudinal field with a characteristic strength of several hundred G, that likely changes sign.

\begin{figure*}
\begin{centering}
\includegraphics[width=7cm,angle=90]{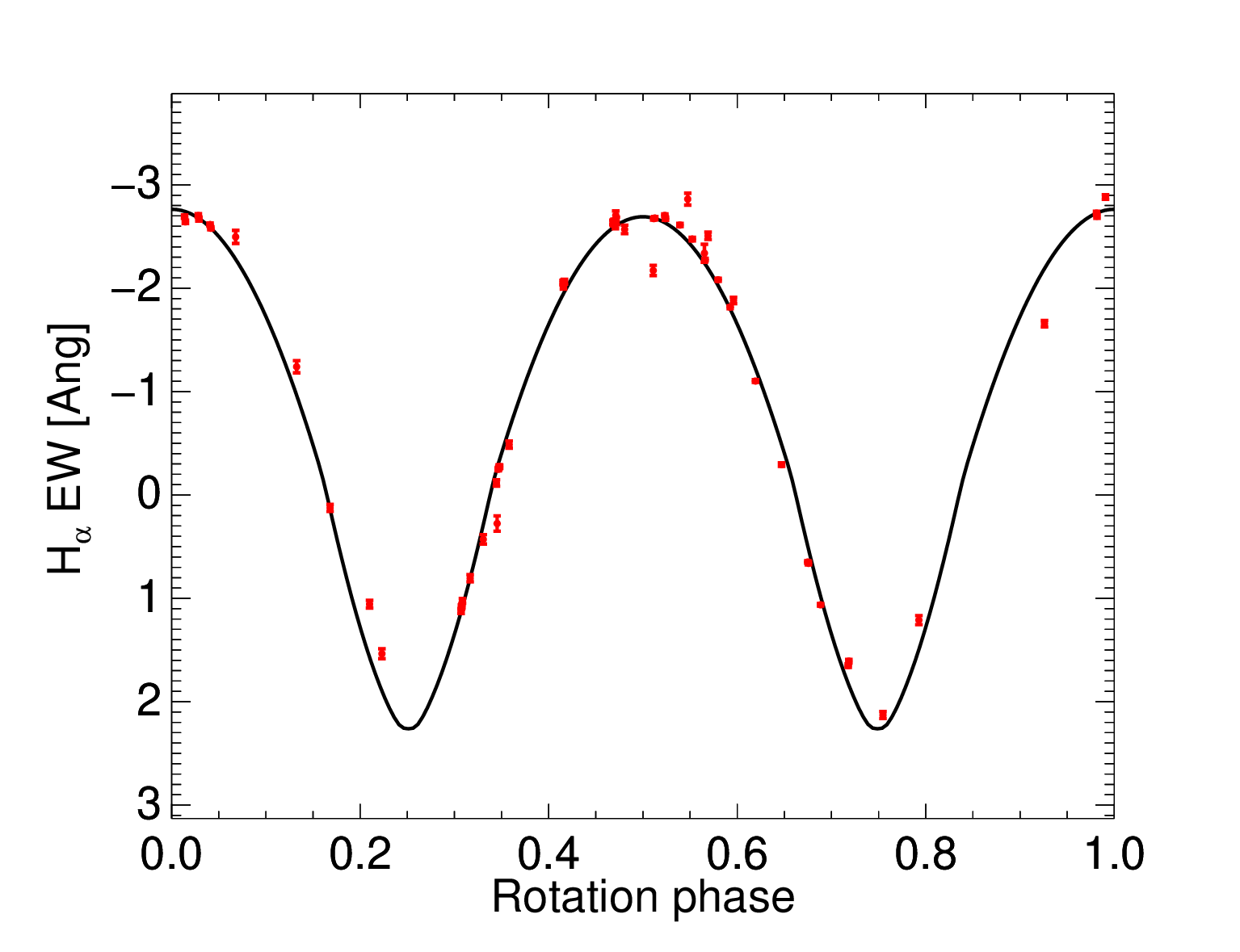}\includegraphics[width=7cm,angle=90]{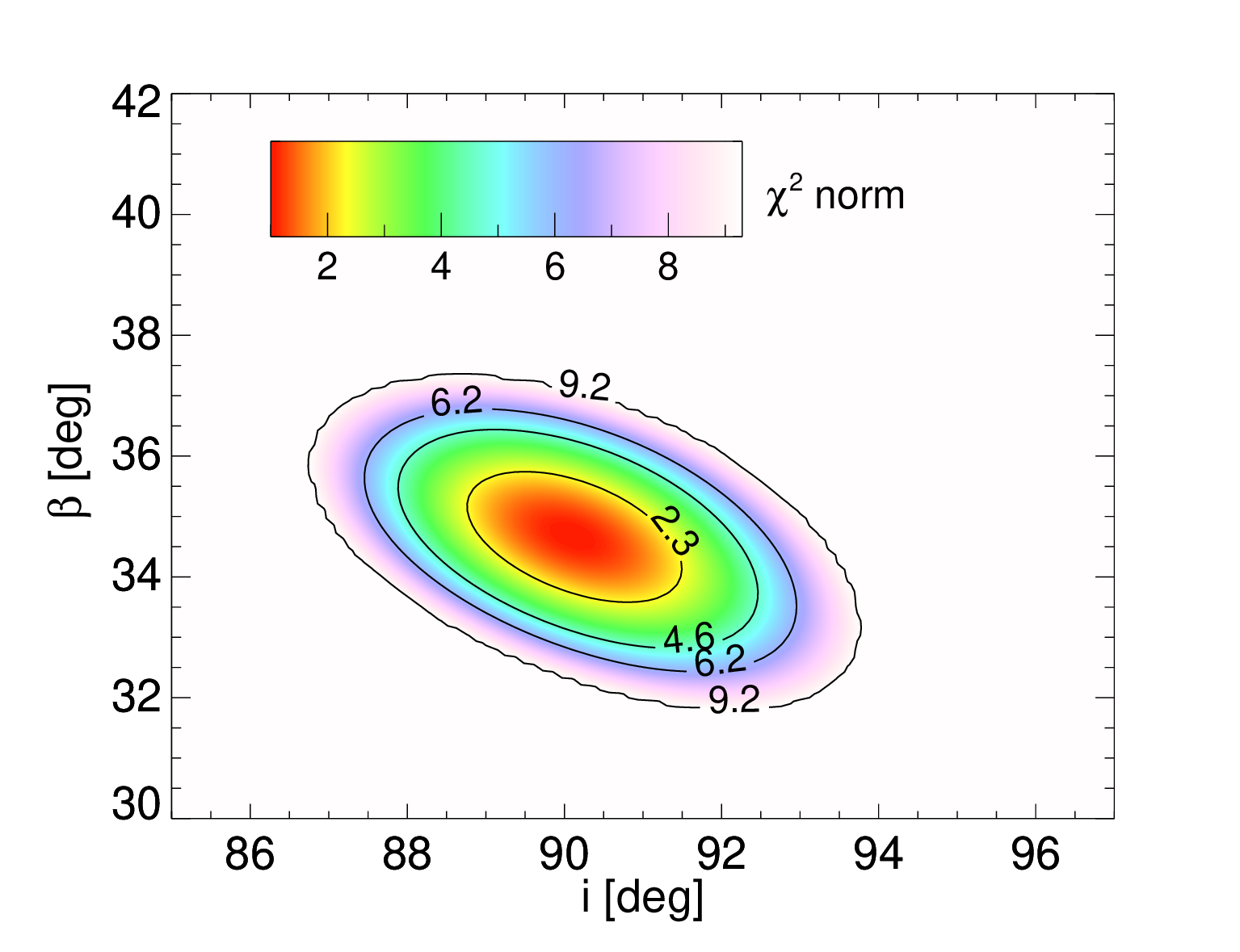}
%\vspace{3in}
\caption{\label{mhd2}MHD modelling of the H$\alpha$ variation.  {\em Left -}\ Model fit to the phased H$\alpha$ EW variation. {\em Right -}\ $\chi^2$ contour map illustrating the best fit solution of the magnetic geometry. See text for details.}
\end{centering}
\end{figure*}

\section{Stellar and magnetic geometry}

With the inferred rotational period and derived radius (from Table 1), \cpd\ should have an equatorial rotational velocity in the range $v_{\rm e}\simeq 7-11$~km\,s$^{-1}$. Since the upper limit on $v\sin i$ obtained in Sect. 3 (see Table 4) is much larger than this value due to the significant turbulent broadening of the line profile, we are not able to constrain the inclination of the stellar rotational axis via comparison of $v_{\rm e}$ and $v\sin i$. 

Instead, we derive both the rotational and magnetic geometry of \cpd\ from considering the rotationally modulated 
H$\alpha$ line stemming from the star's { hypothetical} circumstellar "dynamical 
magnetosphere" \citep{2012MNRAS.tmpL.433S}. Below the Alfven radius $R_{\rm A}$ at which 
the magnetic and wind energy densities are equal, the magnetic field is strong 
enough to channel the radiatively driven wind outflow of \cpd\ along closed field lines (ud-Doula \& Owocki 2002). 
The trapped wind plasma, channeled along field lines from opposite magnetic hemispheres then 
collides at the magnetic equator, and is pulled back to the star's surface by gravity. 
This in turn leads to a statistically overdense region centered 
around the magnetic equator, which is also characterized by infalling material of quite 
low velocities (as compared to non-magnetic O star wind velocities).  If the magnetic 
and rotational axes of the star are mutually inclined, an observer at earth will
view this dynamical magnetosphere from different perspectives, which 
leads to rotationally modulated line profiles \citep{2012MNRAS.tmpL.433S,
2012MNRAS.426.2208G,2013MNRAS.428.2723U,2013MNRAS.429..398P}. Below we use this
variability to derive constraints on the magnetic geometry of \cpd. 
   
We follow the procedure developed by \citet{2012MNRAS.tmpL.433S} (see also \citet{2012MNRAS.426.2208G} and \citet{2013MNRAS.428.2723U}) and use 100 snapshots of a 2-D radiation magnetohydrodynamic (MHD) 
wind simulation of a magnetic O-star, that we patch together in azimuth to form a 3-D "orange slice" 
model. { The MHD simulation employed here is that computed for HD 191612 \citep[see][]{2012MNRAS.tmpL.433S} as a proxy for \cpd.}
This is a reasonable approach because both HD 191612 and \cpd\ are slow rotators (in the sense that rotation is dynamically insignificant 
in determining the wind/magnetosphere properties), and their magnetic field strengths/confinement parameters are
formally identical (see Sect. 7). Hence the geometry of the plasma confinement should be similar in both cases \citep{2002ApJ...576..413U}.

%However, as we discussed with Jason the other week, the exact value of Bp (or, for this purpose, eta_star) is actually not very important for the analysis of the Ha *variation* using this technique. This is because the Alfven radius only goes with eta_star ^ 1/4. This means also a somewhat different magnetic field strength would give essentially the same result (I have actually checked this using another MHD model as input to the RT computations, and it is true). 

The tests by \citet{2013MNRAS.428.2723U} show this patching technique results in a 
reasonably good representation of the full 3-D magnetosphere. To compute synthetic H$\alpha$ 
spectra, we then solve the formal solution of radiative transfer in a 3-D cylindrical system 
for an observer viewing from angle $\alpha$ with respect to the magnetic axis. For 
magnetic obliquity $\beta$ and observer inclination $i$ we then have 
\begin{equation} 
	\cos \alpha = \sin \beta \cos \Phi \sin i + \cos \beta \cos i, 
\end{equation} 
% 
%\noident (e.g. Gagn\'e et al. 2005) 
which gives the observer's viewing angle as function of rotational phase 
$\Phi$, thus mapping out the rotational phase variation for a given 
couple of $\beta$ and $i$\footnote{Although the 2D MHD simulations of \citet{2002ApJ...576..413U} assumed field-aligned rotation,
the slow rotation of \cpd, and hence the lack of any significant dynamical influence of rotation, allows us to construct 3D non-aligned structures
for synthesis of H$\alpha$.}. We solve the formal solution only in the 
wind, assuming H$\alpha$ occupation numbers and a temperature structure 
given by a 1-D NLTE model atmosphere calculation, and using an input photospheric 
H$\alpha$ line profile as lower boundary condition (see \citet{2012MNRAS.tmpL.433S} for 
more details). The infalling material in our selected 100 
snapshots falls preferably toward one pole. This is most presumably due to a 
subtle numerical issue in the MHD simulations, where over a time this north-south asymmetry 
is canceled out \citep[see][]{2002ApJ...576..413U}. To account for 
the fact that in these non-rotating stars there should be no preference between the north and south 
magnetic pole, we here average computed line profiles from the same angle 
$\alpha$ as counted from the north and south magnetic poles, respectively.
  
The absolute level of H$\alpha$ emission should 
further, in principle, provide constraints on the rate by which the magnetosphere is fed by radiatively driven 
wind material (in analogy with how H$\alpha$ emission from non-magnetic 
O-stars provides constraints on the stellar mass loss). However, as discussed in detail by 
\citet{2012MNRAS.426.2208G}, adjusting the underlying model so that the level of H$\alpha$ emission is reproduced 
in the high state results in variability between the high and low states that is too small to reproduce the observations. 
Since here we are mainly interested in obtaining constraints on the magnetic geometry from the 
variability itself, we thus compute synthetic H$\alpha$ equivalent width curves that have somewhat too 
strong absolute emission and then simply shift them down so that the absolute level of emission 
at the extrema are fit. This then results in equivalent width curves that reproduces well the 
variability, providing good constraints on the geometry. However, because of this 
shifting we are not able to simultaneously derive constraints on the mass feeding rate; this 
issue will be discussed in detail in a forthcoming paper.      

%\begin{figure*}
%\begin{centering}
%\includegraphics[width=6.65cm,angle=90]{rho2.eps}\includegraphics[width=6.65cm,angle=90]{vrad.eps}
%%\vspace{3in}
%\caption{\label{mhd1}MHD modelling of the H$\alpha$ variation. {\em Left -}\ Density. {\em Right -}\ radial velocity. {\vero Jon, would you please elaborate.}}
%\end{centering}
%\end{figure*} 

The right panel of Fig.~\ref{mhd2} shows the $\chi^2$ landscape from fitting the observed rotational phase variation for given 
sets of $\beta$ and $i$. Because of the strong emission dependence on $\alpha$, the error bars of the 
best fit $i = 90$ deg and $\beta = 35$ deg are quite small. We note, however, that simply switching the 
obliquity and inclination angles gives equal results, i.e. there is a second "best-fit" model at $i = 35$ deg 
and $\beta = 90$ deg (that is not shown in the figure). The left panel then finally compares the best model with the 
observed variability as function of rotational phase. 

\subsection{Photometric and broadband polarization variability}

A second constraint on the geometry is potentially derived from photometric and broadband (linear) polarimetric variability. We use the Monte-Carlo radiative transfer (RT) code developed by RHDT for simulating light scattering in circumstellar envelopes, first applied in this context to the Of?p star HD 191612 by \citet{2011MNRAS.416.3160W}. In this code, photon packets are launched from a central star and allowed to propagate through an arbitrary distribution of circumstellar matter (described by a Cartesian density grid), until they are scattered by free electrons. Upon scattering, a ray is peeled off from the packet toward a virtual observer, who records the packet's Stokes parameters appropriately attenuated by any intervening material \citep[see][for a discussion of this peel-off technique]{1984ApJ...278..186Y}. A new propagation direction is then chosen based on the dipole phase function \citep{1960ratr.book.....C}, and the packet's Stokes parameters are updated to reflect the linear polarization introduced by the scattering process. The propagation is then resumed until, after possible further scatterings, the packet eventually escapes from the system or is reabsorbed by the star.

\begin{figure*}
\begin{centering}
\includegraphics[width=18cm]{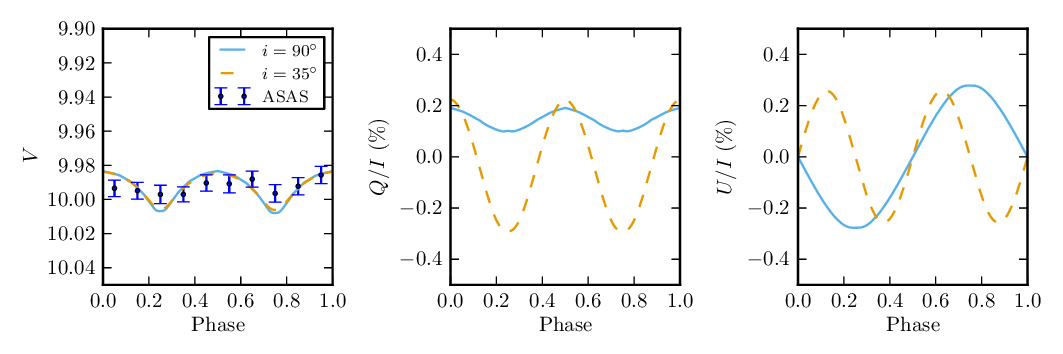}
%\vspace{3in}
\caption{\label{townsend}Predicted light curve and broadband polarization variations of \cpd. Variations are shown for two models: ($i=35\degr, \beta=90\degr$), and ($i=90\degr, \beta=35\degr$). Whereas both models produce identical light curves, their polarization variations are clearly distinct.}
\end{centering}
\end{figure*}

We employ the circumstellar density model developed for HD 191612, and compute lightcurves for the two geometries derived from the orange-slice MHD modelling. In Fig.~\ref{townsend} we compare the observed ASAS $V$-band photometric variation with the predictions of the Monte Carlo RT code. Unfortunately, the available photometry is not sufficiently precise to allow us to obtain meaningful constraints on the geometry. Nevertheless, the predictions demonstrate that the existing photometry is at the threshold of being able to detect the predicted variation. High-precision photometry and broadband polarimetry would provide an additional constraint on the geometry, and in particular the polarimetric variation would enable the determination of the individual values of the angles $i$ and $\beta$.

\subsection{Longitudinal magnetic field variation}

\begin{figure}
\begin{centering}
\includegraphics[width=8cm]{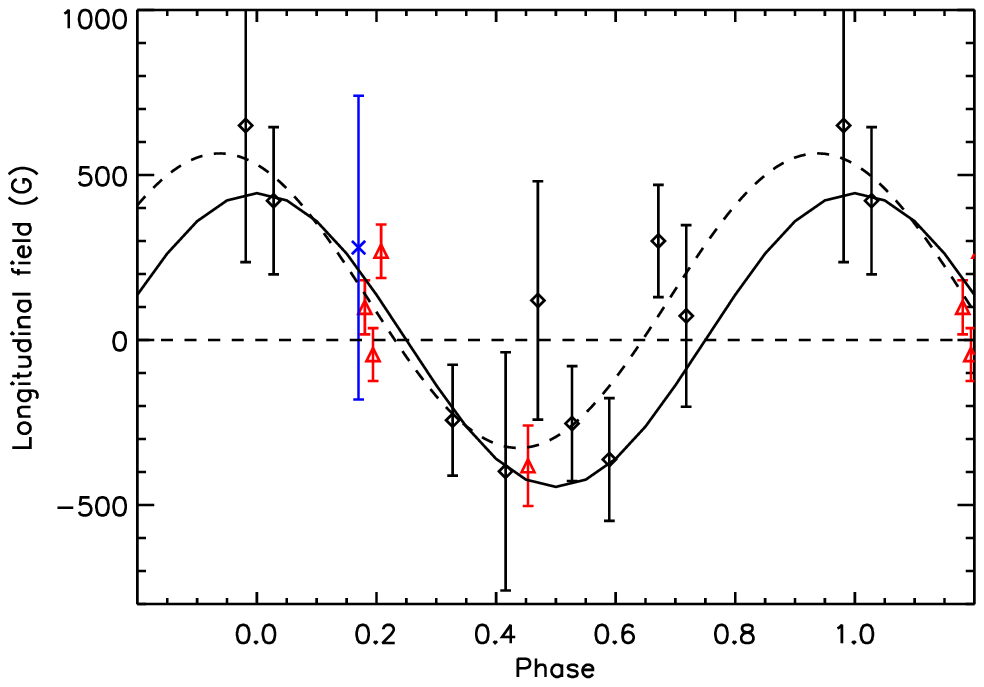}
\includegraphics[width=8cm]{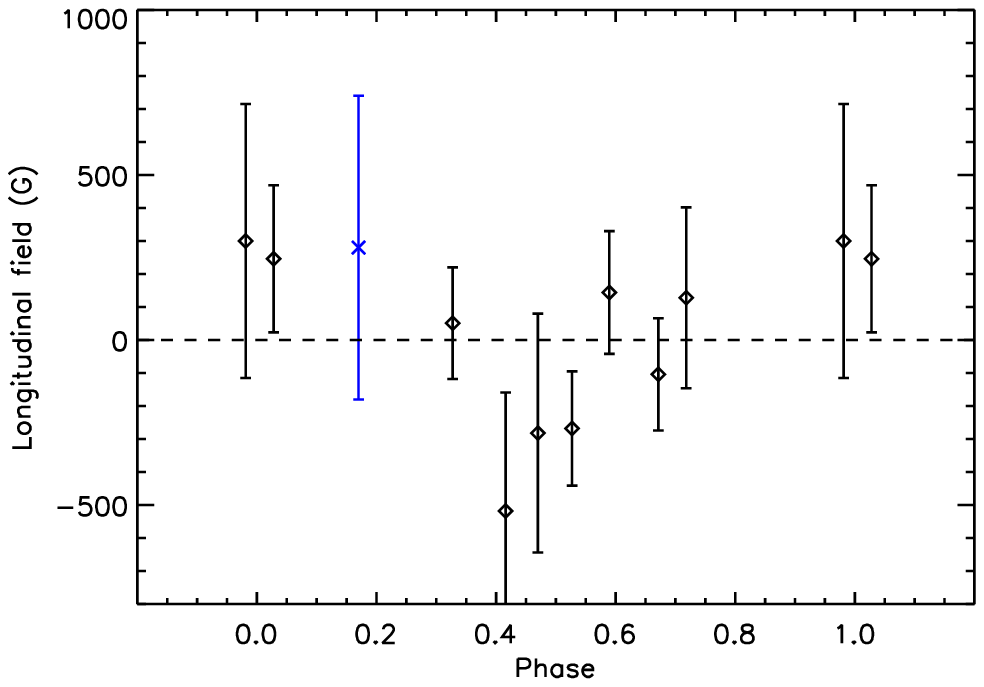}
%\vspace{3in}
\caption{\label{bz}{\em Top panel:}\ Longitudinal field versus phase. Dashed line - best-fit sinusoid. Solid line - Dipole model (fixed phase and geometry, polar strength $B_{\rm d}=2.6$~kG). Diamonds represent ESPaDOnS measurements from the current study. The cross is the HARPSpol measurement. The triangles are measurements from \citet{2011A&A...528A.151H,2013A&A...551A..33H}. {\em Bottom panel:}\ Null field versus phase. All measurements are phased according to Eq. (1).}
\end{centering}
\end{figure}

To infer the strength of the magnetic field, and to test its compatibility the derived geometry, we model the longitudinal field variation of \cpd\ as a function of rotational phase. We used the coadded profiles discussed in Sect. 5 and reported in Table~\ref{tab:specpol}, phased according to Eq. (1). This phase variation $\bz(\phi)$ of the longitudinal field is illustrated in Fig.~\ref{bz} (upper frame). A fit by Least-Squares of a cosine curve of the form $\bz(\phi)=B_{\rm 0}+B_{\rm 1}\cos(2\pi(\phi-\phi_{\rm 0}))$ to the data yields a reduced $\chi^2$ of 0.84, with parameters $B_{\rm 0}=+115\pm 55$~G, $B_{\rm 1}=450\pm 100$~G and $\phi_{\rm 0}=0.18\pm 0.15$. 

Therefore, according to Least Squares, the variation of the field is detected at $4.5\sigma$ confidence.  The reduced $\chi^2$ of the data relative to the straight line $\bz(\phi)=0$ (the hypothesis of a null field) is 2.3. These results indicate that the variation is significant at about 97\% confidence, and that the null hypothesis can be rejected as an acceptable representation of the data at similar confidence. The longitudinal field measured from the null profiles (lower frame of Fig.~\ref{bz}) yields similar reduced $\chi^2$ for both the cosine and straight-line fits, and indicates no significant variation.

Adopting $i=35\degr$ and $\beta=90\degr$ from the modelling of the H$\alpha$ EW, we have fit the phase-binned longitudinal field measurements with a synthetic longitudinal field variation with fixed phase of maximum (phase 0.5) and variable polar field strength $B_{\rm d}$. The best-fit model (according to the $\chi^2$ statistic) is characterised by $B_{\rm d}\simeq 2.6$~kG.  The extrema of the best-fit dipole model are slightly offset from the best-fit sinusoid in mean longitudinal field strength (by about 100 G) and in phase (by about 0.07 cycles). These offsets are well within the uncertainties of the observed variation. Taking them into account, we estimate an uncertainty on the derived dipole strength of $\pm 900$~G. 

If we fix only the inclination and allow both $\beta$ and $B_{\rm d}$ to vary, a direct fit to the longitudinal field variation yields best-fit values of $\beta=78^{+10}_{-8}\degr$ and $B_{\rm d}=2.7\pm 1.1$~kG. These values are in good agreement with those derived from the fit with fixed geometry derived above.

In Fig.~\ref{bz} we also show the longitudinal field measurements of Hubrig et al. (2011, 2013) obtained from 'all' lines. Those measurements, which have formal errors that are substantially more precise than our own, are in good agreement with both the best-fit synthetic dipole and the best sinusoidal fit. It is notable that 3 of the measurements of Hubrig et al. were acquired at essentially the same phase, which corresponds to crossover (i.e. $\bz\simeq 0$).

%\begin{figure}
%\begin{centering}
%\subfloat[][Longitudinal field versus phase. Dashed line - best-fit sinusoid. Solid line - Dipole model (fixed phase and geometry, polar strength $B_{\rm d}=2.6$~kG.) Diamonds represent measurements from the current study. Triangles are measurements from Hubrig et al. (2011, 2013).]{\includegraphics[width=8cm]{cpd-282561_bz.ps}}
%
%\subfloat[][Null field versus phase.]{\includegraphics[width=8cm]{cpd-282561_nz.ps}}
%%\vspace{3in}
%\caption{\label{bz}Longitudinal magnetic field variations from Stokes $V$ and diagnostic null, phased according to Eq. (1).}
%\end{centering}
%\end{figure} 

\subsection{Modeling the Stokes $V$ profiles}

We also modelled the magnetic field geometry using the LSD Stokes $V$ profiles. We compared the mean LSD profiles for phases 0.0, 0.5 and 0.75 to a grid of synthetic Stokes $V$ profiles using the method of Petit \& Wade (2012). For this modelling we used LSD profiles extracted using a metallic line mask, as described by \citet{2012MNRAS.419.2459W}.

The emergent intensity at each point on the stellar surface is calculated using the weak-field approximation for a Milne-Eddington atmosphere model. In this model, the source function is linear in optical depth such that $S(\tau_c)=S_0[1+\beta\tau_c]$. We use $\beta=1.5$, Voigt-shaped line profiles with a damping constant $a=10^{-3}$ and a thermal speed $v_\mathrm{th}=5$\,km\,s$^{-1}$. 
The line-to-continuum opacity ratio $\kappa$ is chosen to fit the intensity LSD profile. 
The synthetic flux profiles are then obtained by numerically integrating the emergent intensities over the projected stellar disk. The projected rotational velocity is set to $v\sin i=9$\,km\,s$^{-1}$. We applied isotropic Gaussian macroturbulence\footnote{of the form $\mathrm{e}^{-v^2/v^2_\mathrm{mac}}/(\sqrt{\upi}v_\mathrm{mac})$} compatible with that determined in Sect. 3.

We assume a simple centred dipolar field, parametrized by the dipole field strength $B_d$, the rotation axis inclination $i$ with respect to the line of sight, the positive magnetic axis obliquity $\beta$ and the rotational phase $\varphi$. 
Assuming that only $\varphi$ may change between different observations of the star, the goodness-of-fit of a given rotation-independent ($B_d$, $i$,	$\beta$)	magnetic configuration can be computed to determine configurations that provide good posterior probabilities for all the observed Stokes $V$ profiles in a Bayesian statistic framework. 
In order to stay general, we do not at this point constrain the rotational phases of the observations nor the inclination of the rotational axis. 

The Bayesian prior for the inclination is described by a random orientation $[p(i) = \sin(i)\,di]$, the prior for the dipolar field strength has a modified Jeffreys shape to avoid a singularity at $B_d = 0$\,G, and the obliquity and the phases have flat priors. 

To assess the presence of a dipole-like signal in our observations, we compute the odds ratio of the dipole model ($M_1$) with the null model ($M_0$; no magnetic field implying Stokes $V = 0$). We also perform the same analysis on the null profiles. The results are displayed in Table\,\ref{tab:raven_odds}.
Taking into account all the observations simultaneously, the odds ratio is in favour of the magnetic model by 9 orders of magnitude. 
For the null profiles, the combined odds ratio is 2:1 in favour of the null model. 
Note that as the case $B_d = 0$\,G is included in the magnetic model, in the latter case the difference between the two models, which can equally well reproduce a signal consisting of only pure noise, is expected to be dominated by the ratio of priors in this case, i.e. the Occam factor that penalizes the magnetic model for its extra complexity. 

\begin{figure}
%\begin{centering}
%\hspace{-0.75cm}\includegraphics[width=0.4\textwidth,angle=90]{CPD-28-2561-post-tri.ps}
\includegraphics[width=0.4\textwidth]{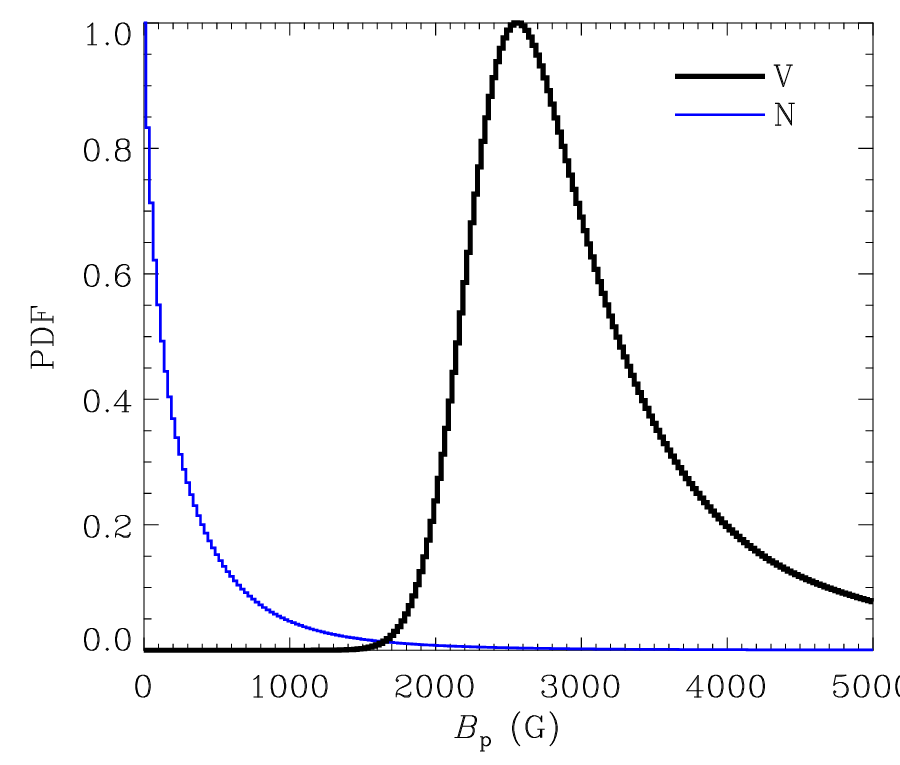}
\includegraphics[width=0.4\textwidth]{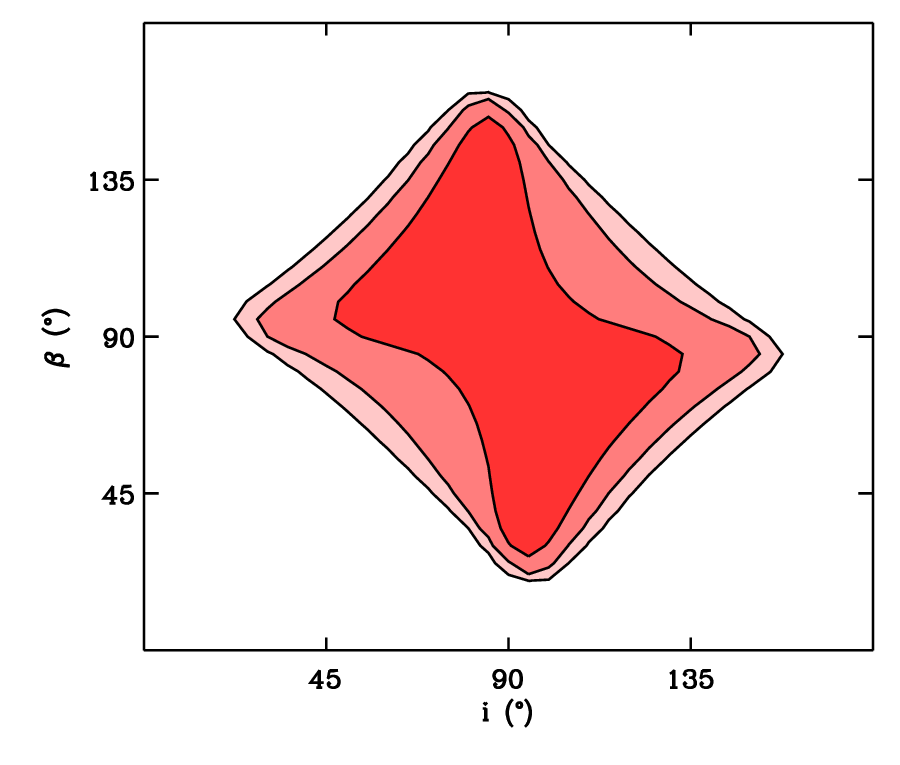}
\caption{\label{fig:raven}Magnetic field polar strength (upper frame) and geometry $(i,\beta)$ (lower frame) constraints derived from of modelling of Stokes $V$ profiles.}
%\end{centering}
\end{figure}

Figure~\ref{fig:raven} shows the posterior probability density function for each model parameters. The 68.3, 95.4, 99.0, and 99.7 percent regions, tabulated in Table\,\ref{tab:raven_reg}, are illustrated in dark to pale shades, respectively.  At 95.4\% confidence, the polar strength of the dipole magnetic field of \cpd\ is found to be in the range $1.9\leq B_{\rm d}\leq 4.5$~kG. This is in good agreement with the dipole strengths derived from the longitudinal field variation.

%Note for Vero
%{Here I would comment on the sharp lower edge of the probability, and perhaps how the geometry matches was was inferred in the previous sections. As I said, I can re-do a calculation later with a gaussian prior for the inclination, centred on the value determined from the period/radius.}

%------
\begin{table}
	\caption{\label{tab:raven_odds}Odds ratios derived from the analysis of Stokes $V$ profiles.}
	\begin{center}
	\begin{tabular}{l  c c}
	\hline
Phase& $\log(M_0/M_1)$ $V$ & $\log(M_0/M_1)$  $N$ \\
\hline
0.00 &   -0.91  &       0.09 \\
0.50 &      -7.43  &        0.24 \\
0.75 &    -0.34  &       0.06 \\
\hline
Combined &       -9.06   &   0.25 \\
\hline
	\end{tabular}
	\end{center}
\end{table}
%----------

%------
\begin{table}
	\caption{\label{tab:raven_reg}Credible regions derived from the Bayesian analysis of Stokes $V$ profiles.}
	\begin{center}
	\begin{tabular}{l  c}
	\hline
 Credible  & Range in gauss \\	
 region & \\
\hline
	\multicolumn{2}{c}{$V$} \\
   99.7  &  1754  -   5000 \\
   99.0  &  1841  -   4947 \\
   94.5  &  1891  -   4509 \\
   68.3  &  2137  -   3307 \\
	\hline
	\multicolumn{2}{c}{$N$} \\
    99.7 &  0  -  3835 \\
    99.0 &  0  -  2767 \\
    95.4 &  0  -  1508 \\
    68.3 &  0  -   425 \\
   \hline
	\end{tabular}
	\end{center}
\end{table}
%----------

\section{Magnetosphere}

%\vero{Gregg, would you arrange the next two paragraphs to make sure we are not repeating?}

%Stellar magnetospheres form through the channelling and confinement of an outflowing wind by the star's magnetic field. 
%This magnetic control breaks the symmetry of radiatively-driven winds, and therefore will influence observable wind diagnostics, such as Balmer line emission \citep{1978ApJ...224L...5L} and wind resonance lines in the UV \citep{1990ApJ...365..665S}. Furthermore, material forced to flow along the field lines will collide near the tops of closed loops, producing a shock-heated volume of plasma that will eventually cool, radiating X-rays \citep{1997ApJ...485L..29B,1997A&A...323..121B}. Given the large mass-loss rate expected for a 35\,kK O-type star, and the observed strong magnetic field, it is highly probably that such a structure exists around CPD-28\,2561.

As presented by \citet{2002ApJ...576..413U}, the global competition between the magnetic field and stellar wind can be characterized by the so-called wind magnetic confinement parameter $\eta_\star \equiv B^2_\mathrm{eq}R^2_\star / \dot{M}_{\rm B=0}v_\infty$, 
which depends on the star's equatorial field strength ($B_\mathrm{eq}$), stellar radius ($R_\star$), and wind momentum ($\dot{M}_{\rm B=0}v_\infty$) the star would have in absence of the magnetic field. For a dipolar field, one can identify an Alfv\'en radius $R_\mathrm{A}\simeq\eta_\star^{1/4}R_\star$, representing the extent of strong magnetic confinement. 
Above $R_\mathrm{A}$, the wind dominates and stretches open all field lines. But below $R_\mathrm{A}$, the wind material is trapped by closed field line loops, and in the absence of significant stellar rotation is pulled by gravity back onto the star within a dynamical (free-fall) time-scale.

To estimate the Alfv\'en radius of the magnetosphere of \cpd, we use the stellar parameters given in Table 4. 
% Note for Vero
%BP_arr = [2137, 2650, 3307]
%Mdot=1e-6 Msun/yr
%vinf=2400 km/s (assumed by Fabrice)
%vrot=9 km/s (R=12.9 Rsun, Prot=73.4d)
%ms_arr = [60., 35., 30.]
%rs_arr = [10., 13., 16.]

\begin{figure}
	\includegraphics[width=0.45\textwidth]{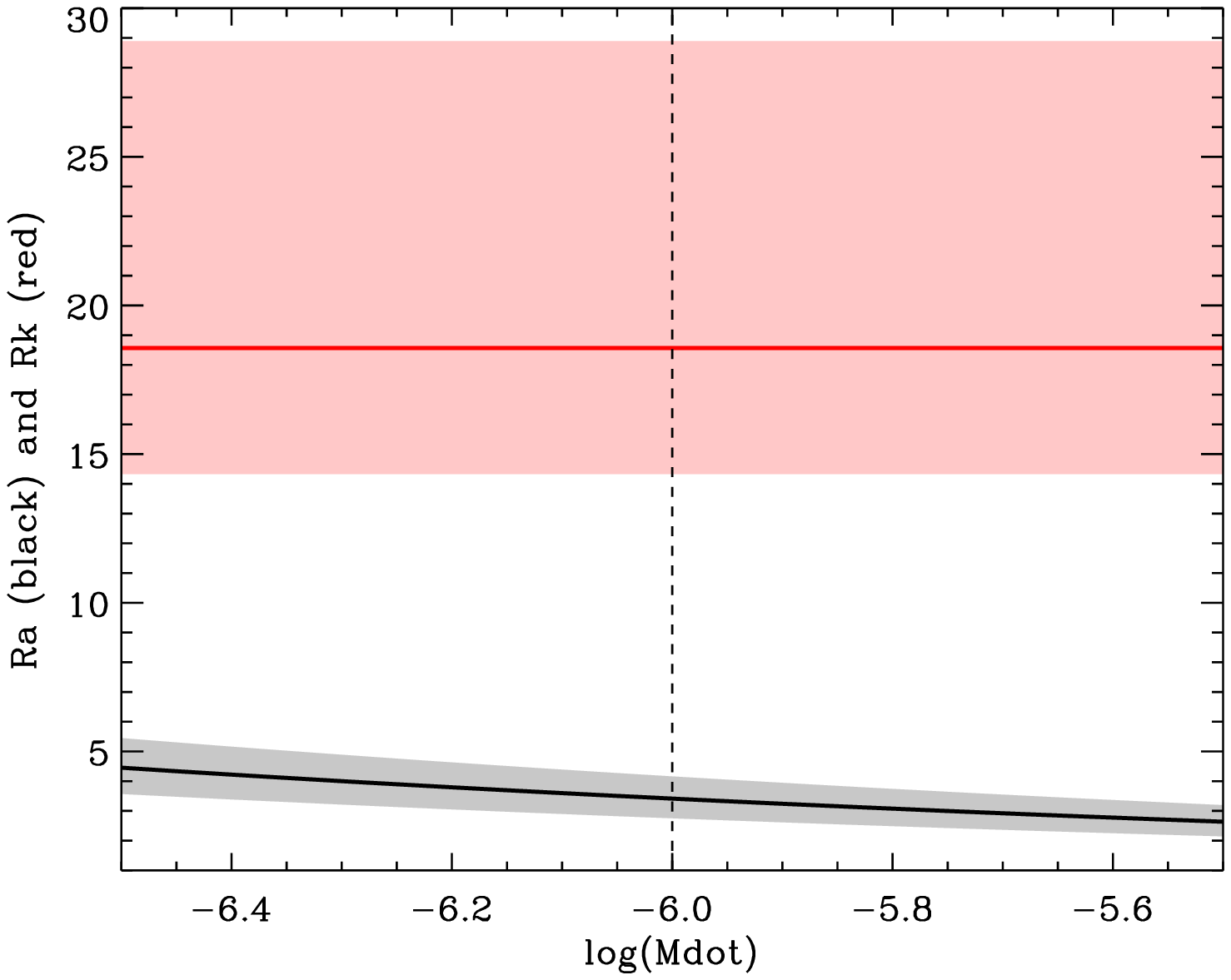}
	\caption{\label{fig:magneto} Alfven (black, lower curve) and Kepler (red, upper curve) radii as a function of the mass-loss rate.}
\end{figure}

The stellar parameter with the largest uncertainty is the wind momentum. Fig.\,\ref{fig:magneto} therefore illustrates the variation of $R_\mathrm{A}$ with one order of magnitude variation in mass-loss rate (corresponding to a generous estimate of the uncertainty). One can see how this uncertainty is mitigated by the 1/4 power dependence of the wind momentum in the definition of the Alfven radius. 
The grey shaded areas represent intervals of stellar radius uncertainty ($\pm3$\,R$_\odot$) and of dipole field strength (2.1, 2.6, 3.3 kG, reflecting the 68.3\% Bayesian credible region) meant to minimize and maximize the Alfven radius.

We therefore expect the Alfven radius to be of the order of 3 stellar radii, and certainly no more than 5 stellar radii. 

In the presence of significant stellar rotation, centrifugal forces can support any trapped material above the Kepler co-rotation radius $R_\mathrm{K}\equiv(GM/\omega^2)^{1/3}$. This requires that the magnetic confinement extend beyond this Kepler radius, in which case material can accumulate to form a {\em centrifugal magnetosphere} \citep[e.g.][]{2005ApJ...630L..81T}. 

In the case of \cpd, the slow rotation puts the Kepler radius much farther out than the Alfven radius ($\sim18$\,R$_\star$) and no long-term accumulation of wind plasma is expected, as illustrated by the red curve in Figure\,\ref{fig:magneto} (the red shaded area represents $\pm3$\,R$_\odot$ and a range of mass from 30 to 60\,M$_\odot$). In such a {\em dynamical magnetosphere} configuration, transient suspension of circumstellar material results in a statistical global over-density in the closed loops. For O-type stars with sufficient mass-loss rates, the resulting dynamical magnetosphere can therefore exhibit strong emission in Balmer recombination lines \citep{2012MNRAS.tmpL.433S,2013MNRAS.429..398P}. This conclusion supports the MHD modelling employed to determine the stellar geometry in Sect. 6. We note that due to infalling wind material { (which may be reflected in the P Cyg-like profile of He~{\sc ii} $\lambda 4686$ at some phases)}, the global mass-loss rate of a star exhibiting such a dynamical magnetosphere is significantly reduced. According to the scaling relations of \citet{2008MNRAS.385...97U} (their eqn. 10 and 23), for an estimated $r_{\rm A} \approx 3 R_{\rm star}$, $\dot{M}/\dot{M}_{B =0} \approx 0.2$, i.e. the global mass-loss rate is reduced by approximately a factor of 5 due to wind plasma falling back upon the star. 

%\vero{Gregg, the following is probably not needed anymore. However, it seems a bit strange to me to have the magnetospheric calculations come in after Jon's Halpha models. Perhaps these two sections should be merged together?}
%However, the transient suspension of circumstellar material still results in a global over-density in the closed loops. For O-type stars with sufficient mass-loss rates, the resulting dynamical magnetosphere can therefore exhibit strong emission in Balmer recombination lines \citep{2013MNRAS.433.2497S,2013MNRAS.429..398P}. 

The characteristic magnetic braking time can be estimated using equation 25 of \cite{2009MNRAS.392.1022U}. 
Using the nominal wind parameters in Table~\ref{params} and $k\sim0.1$ \citep{2004A&A...424..919C}, we obtain a spin-down timescale of 0.45\,Myr. However as noted by Petit et al. (2013), the square-root dependence of this quantity on the wind momentum renders spin-down estimates valid only to a factor of a few. 

% Gregg, here are the magnetospheric values with the nominal stellar parameters. 
%         Vorb mass: 719.63      km/s
%         Vorb logg: 947.20      km/s
%              Vrot: 8.89      km/s
%            W mass: 0.01          
%            W logg: 0.01          
%                Rk: 18.71        Rs
%             efold: 4.39          
%         omega obs: 0.00          
%        omega crit: 0.00          
%             Omega: 0.02          
%             Re/Rp: 1.00          
%            B crit: 137.21     Gauss
%          eta star: 93.25          
%                Ra: 3.40        Rs
%        log(Ra/Rk): -0.74          
%              taum: 35.00       Myr
%              tauj: 0.45       Myr
%                ts: 1.99       Myr

\section{Discussion and conclusions}

In this paper we have performed a first thorough analysis of the variability, geometry, magnetic field and wind confinement of the Of?p star \cpd. Using more than 75 new medium and high resolution spectra, we determined the equivalent width variations and examined the dynamic spectra of photospheric and wind-sensitive spectral lines, deriving a rotational period of $73.41$~d. We confirmed the detection of an organized magnetic field via Zeeman signatures in LSD Stokes $V$ profiles. The phased longitudinal field data exhibit a weak sinusoidal variation, with maximum of about 565 G and a minimum of about -335 G, with extrema approximately in phase with the H$\alpha$ equivalent width variation. Modeling of the H$\alpha$ equivalent width variation assuming a 3D 'orange-slice' magnetospheric model yields a unique solution for the ambiguous couplet of inclination and magnetic obliquity angles: $(i, \beta)$ or $(\beta, i)=(35\degr,90\degr)$. Assuming this geometry, the surface magnetic field dipole strength is inferred to be $B_{\rm d}=2.6\pm 0.9$~kG. The magnetic strength and rotational period of \cpd\ are rather typical for magnetic O-type stars.

Using the magnetic field strength and inferred wind properties, we derive the wind magnetic confinement parameter $\eta_*\simeq 93$, yielding an Alfv\'en radius $R_{\rm A}\simeq 3.4~R_*$, and a Kepler radius $R_{\rm K}\simeq 18.7~R_*$. This supports a picture in which the H$\alpha$ emission and other line variability have their origin in an oblique, co-rotating 'dynamical magnetosphere' structure resulting from a magnetically channeled wind. This framework is consistent with that inferred for the other Galactic Of?p stars HD 108 \citep{2010MNRAS.407.1423M}, HD 148937 \citep{2012MNRAS.419.2459W}, HD 191612 \citep{2006MNRAS.365L...6D,2011MNRAS.416.3160W}, and NGC 1624-2 \citep{2012MNRAS.425.1278W} as well as the cooler magnetic O-type star HD 57682 \citep{2012MNRAS.426.2208G}. The computed spindown time of \cpd\ (equal to 0.45 Myr) is similar to that of HD 191612 (0.33 Myr). On the other hand, it is significantly shorter than that of HD 148937 \citep{2012MNRAS.419.2459W}. At face value it is also significantly shorter than that of HD 108 \citep[1-3 Myr][]{2010MNRAS.407.1423M}, however the value for HD 108 is based on a highly uncertain estimate of the polar field strength. Considering the large range of rotational periods of these stars, and the general lack of accurate information concerning their ages, it is as yet difficult to derive any meaningful conclusions about the role of magnetic braking in the evolution of these stars.

Most lines in the spectrum of \cpd\ are detectably variable, including lines of H, He~{\sc i}/{\sc ii}, and (more weakly) C~{\sc iv}. The equivalent width variation of most lines is "double-wave", exhibiting two maxima (and minima) per rotational cycle. The EW variability and extrema of most lines are roughly symmetric in both phase and amplitude, although the maximum/minimum at phases 0.25/0.5 are somewhat (10-20\%) less pronounced in some lines than those at phases 0.75/0.0. The He~{\sc ii}~$\lambda 4542$ line is the only line investigated that exhibits apparently "single-wave" variability. The double-wave nature of the variations of most lines can be naturally explained as the consequence of the stellar and magnetic geometry of \cpd : the derived inclination and obliquity angles yield $i+\beta=135\degr$, implying that both magnetic poles are presented clearly to the observer during each stellar rotation. Only two other magnetic O-type star are known to exhibit analogous behaviour: HD~57682 \citep{2012MNRAS.426.2208G}, for which $i+\beta=139\degr$, and Plaskett's star \citep[e.g.][and Grunhut et al. 2015, in prep.]{2013EAS....64...67G}, for which $i+\beta\simeq 130\degr$. 

The behaviour of the $\lambda 4542$ line can likely be understood as due to the depth of formation and the relative contribution of the magnetosphere to the line profile. \citet{2014arXiv1409.5057M} constrained the line formation region of various lines in spectra of non-magnetic O stars, and investigated the variability as a function of depth. They demonstrate that the Balmer lines are formed further out in the wind compared to some weak He~{\sc i} or high ionization He~{\sc ii} or C~{\sc iv} lines. { Lines formed closer to the photosphere also exhibit some additional variability that likely arises in the photosphere (and this may qualitatively explain the different behaviour of lines such as He~{\sc ii}~$\lambda 4542$). However, the variability of these photospheric lines is weak in comparison to the variability of wind-dominated lines, and cannot contribute appreciably to the variability of features such as H$\alpha$ or He~{\sc ii}~$\lambda 4686$.  We therefore propose that in the case of \cpd\ (and likely other magnetic O stars), lines clearly formed out in the magnetosphere (H$\alpha$, H$\beta$, He~{\sc ii} $\lambda 4686$) are overwhelmingly dominated by the rotational variation of the density of the confined wind plasma. }

The dynamic spectra of \cpd\ reveal characteristic differences in the mean velocities and skewness of the two emission maxima per rotation cycle. This phenomenon - first discussed by \citet{2012MNRAS.426.2208G} in the context of HD~57682 - is not reproduced by the orange-slice MHD model. In the case of HD~57682, the shift between the two emission peaks is about 70~km/s; for \cpd, it is comparable (about 60~km/s). This effect could potentially be explained by a difference in strength and/or geometry of the two magnetic poles at the stellar surface. Such differences are commonly inferred for cooler magnetic stars and are frequently parametrized using a dipole model offset from the stellar centre along its axis. Typically, dipole offsets measured for Ap/Bp stars are 0.0-0.3~$R_*$ \citep[e.g.][]{1997MNRAS.292..748W,2000A&A...355.1080W}. First 2D MHD simulations computed using reasonable dipole offsets provide promising results. However, their presentation and discussion is outside the scope of this paper. This potential ability of dipole offset to explain the observed { emission line variations} will be the topic of a forthcoming paper.

The absolute level of H$\alpha$ emission from the magnetosphere of an O star should provide constraints on the rate by which it is fed by radiatively driven 
wind material through comparison with theoretical models such as those described in Sect. 6. However, as discussed in detail by 
\citet{2012MNRAS.426.2208G}, adjusting the underlying MHD model so that the level of H$\alpha$ emission is reproduced 
in the high state results in variability between the high and low states that is too small to reproduce the observations. This was interpreted by those authors as a consequence of a magnetosphere consisting of somewhat optically thicker plasma 'snakes' (dense plasma clumps falling back toward the star) than predicted by 2D MHD models. Such thicker clumps would lead to higher variability at a given mass-loss feeding rate, and so to a better fit to the observations for the mass-loss feeding rate that reproduces the mean level of emission in the high state.  
%This looks good, except perhaps the part regarding the interpretation of the too small variability in the Grunhut et al. paper. I don?t recall exactly what Jason wrote in that paper, but it is actually not the mass feeding rate that is the main issue for the Halpha variation (the mass feeding issue comes into play when we compare to the weaker-than-expected UV line profiles).  
%
%I?d rather say that for a mass-loss feeding rate that well reproduces the *amount* of emission in the high state, the predicted variability between this low and the high state is somewhat too low. This suggests the magnetosphere consists of somewhat optically thicker ?snakes? (clumps) than predicted by current MHD models. Such thicker clumps would lead to higher variability at a given mass-loss feeding rate, and so to a better fit to the observations for the mass-loss feeding rate that reproduces the *amount* of emission in the high state.  
%
%But as discussed earlier, I think this issue is better discussed in a future paper that focuses more on deriving mass-losss feeding rates 
%of magnetic O-stars from phase-resolved optical+UV observations. So for now, I?d suggest simply dropping the last line in the first paragraph. 
%
%Jon 

For \cpd, we reproduced the EW variation of the H$\alpha$ line by computing synthetic H$\alpha$ equivalent width curves with somewhat too strong absolute emission and then artificially shifting them so that the absolute levels of emission 
at the extrema were fit. While this procedure resulted in equivalent width curves that reproduced well the 
variability, we were not able to simultaneously derive constraints on the mass feeding rate. Recent monitoring of \cpd\ has been performed with HST/STIS (UV spectroscopy) and XMM (X-ray spectroscopy) { [PI: Naz\'e]}. These data will likely help to mitigate the problem of the star's uncertain wind properties, and will provide important insights into its wind confinement and magnetosphere.

In conclusion, we underscore that the weak longitudinal magnetic fields, complex spectra and faint apparent magnitudes of all Of?p stars challenge the capabilities of current polarimetric instrumentation. While their magnetic fields are securely detected, field topologies are only very roughly characterized. The subject of the present paper - \cpd - along with HD 148937, represent the most challenging examples of the class. The development of spectropolarimetric capabilities such as those of ESPaDOnS, but on 8m-class telescopes, is an vital step required in order to investigate stars such as these at an acceptable level of detail, and more generally to extend our understanding of the magnetic properties of the most massive stars.

\section*{Acknowledgments}
GAW acknowledges support from the Natural Science and Engineering Research Council of Canada (NSERC). RHB and JIA acknowledge financial support from FONDECYT, Regular 1140076 and Iniciaci\'on 11121550, respectively. JMA acknowledges support from [a] the Spanish Ministerio de Ciencia e Innovaci\'on through grants AYA2010-15081 and AYA2010-17631; [b] the Junta de Andaluc\'{\i}a grant P08-TIC-4075; and [c] the George P. and Cynthia Woods Mitchell Institute for Fundamental Physics and Astronomy. YN acknowledges support from the Fonds National de la Recherche Scientifique (Belgium), the Communaut\'e Fran\c caise de Belgique, the PRODEX XMM and Integral contracts, and the 'Action de Recherche Concert\'ee' (CFWB-Acad\'emie Wallonie Europe). STScI is operated by AURA, Inc., under NASA contract NAS5-26555. JIA, RB, RG, JMA, AS were Visiting Astronomers, LCO, Chile. JIA and RB were Visiting Astronomers, ESO La Silla, Chile. RG was a Visiting Astronomer, CASLEO, Chile. RHDT acknowledges support from NASA award NNX12AC72G. AuD acknowledges the support by NASA through Chandra Award number TM4-15001A. CFHT and TBL observations were acquired thanks for generous allocations of observing time within the context of the MiMeS Large Programs. The authors acknowledge the contribution by an anonymous referee, whose comments have helped to improve this paper.

\bibliography{cpd}

\clearpage

\end{document}